\documentclass{IEEEtaes}

\usepackage{xcolor}
\usepackage{graphicx}

\usepackage{amsmath,amssymb,amsfonts,mathrsfs,mathtools}
\usepackage{bm}
\usepackage{siunitx}

\usepackage{tabularx,array,makecell,multirow,diagbox}
\usepackage{booktabs}
\usepackage{threeparttable}
\usepackage{afterpage}

\usepackage{subcaption}

\usepackage{algorithm}
\usepackage[noend]{algpseudocode}

\usepackage{url}
\usepackage{lipsum}
\usepackage{multicol}
\usepackage{textcomp}
\usepackage{setspace}
\usepackage{etoolbox}
\usepackage{orcidlink}

\usepackage[backend=biber, style=ieee, maxnames=6, minnames=1, related=false]{biblatex}
\AtBeginEnvironment{procedure}{\let\c@algocf\c@procedure}

\usepackage{hyperref}
\hypersetup{
    colorlinks=true,
    linkcolor=blue,
    citecolor=blue,
    filecolor=blue,
    urlcolor=blue
}

\jvol{XX}
\jnum{XX}
\jmonth{XXXXX}
\paper{1234567}
\pubyear{2025}
\doiinfo{TAES.2025.Doi Number}

\begin{document}

\title{Reliable Transmission of LTP Using Reinforcement Learning-Based Adaptive FEC}

\author{Liang~Chen*\orcidlink{0009-0007-4595-8473}}
\affil{Nanjing University, Nanjing, China} 

\author{Yu~Song*\orcidlink{0009-0008-1064-5909}}
\affil{Eidgenössische Technische Hochschule Zürich, Zürich, Switzerland} 

\author{Kanglian~Zhao\orcidlink{0000-0002-9931-2893}}
\member{Member, IEEE}
\affil{Nanjing University, Nanjing, China} 

\author{Juan~A.~Fraire\orcidlink{0000-0001-9816-6989}}
\member{Senior Member, IEEE}
\affil{Inria, INSA Lyon, CITI, UR3720, 69621 Villeurbanne, France, and CONICET - Universidad Nacional de Cordoba, Cordoba, Argentina}

\author{Wenfeng~Li\orcidlink{0000-0003-2352-3755}}
\member{Member, IEEE}
\affil{Nanjing University, Nanjing, China} 

\receiveddate{Manuscript received XXXXX 00, 0000; revised XXXXX 00, 0000; accepted XXXXX 00, 0000.\\
This work is supported by the National Natural Science Foundation of China (NSFC) under Grant 62131012.}

\corresp{{\itshape (Corresponding author: Kanglian Zhao.)}.\\
*Authors contribute equally to this work.}

\authoraddress{
Liang Chen, Kanglian Zhao, and Wenfeng Li are with the School of Electronic Science and Engineering, Nanjing University, Nanjing 210046, China 
(e-mail: \href{mailto:liangchen@smail.nju.edu.cn}{liangchen@smail.nju.edu.cn}, \href{mailto:zhaokanglian@nju.edu.cn}{zhaokanglian@nju.edu.cn}, \href{mailto:leewf$\_$cn@hotmail.com}{leewf$\_$cn@hotmail.com}).
Yu Song was with Nanjing University, Nanjing 210046, China. He is now with Eidgenössische Technische Hochschule Zürich, Zürich, Switzerland (e-mail: \href{mailto:yusong1@student.ethz.ch}{yusong1@student.ethz.ch}).
Juan~A.~Fraire is with Inria, INSA Lyon, CITI, UR3720, 69621 Villeurbanne, France, and CONICET - Universidad Nacional de Cordoba, Cordoba, Argentina
(e-mail: \href{mailto:juan.fraire@inria.fr}{juan.fraire@inria.fr}).}

\markboth{SONG, CHEN ET AL.}{RELIABLE TRANSMISSION OF LTP USING REINFORCEMENT LEARNING-BASED ADAPTIVE FEC}

\maketitle

\begin{abstract}
Delay/Disruption Tolerant Networking (DTN) employs the Licklider Transmission Protocol (LTP) with Automatic Repeat reQuest (ARQ) for reliable data delivery in challenging interplanetary networks. While previous studies have integrated packet-level Forward Erasure Correction (FEC) into LTP to reduce retransmission time costs, existing static and delay-feedback-based dynamic coding methods struggle with highly variable and unpredictable deep space channel conditions. This paper proposes a reinforcement learning (RL)-based adaptive FEC algorithm to address these limitations. The algorithm utilizes historical feedback and system state to predict future channel conditions and proactively adjust the code rate. This approach aims to anticipate channel quality degradation, thereby preventing decoding failures and subsequent LTP retransmissions and improving coding efficiency by minimizing redundancy during favorable channel conditions. Performance evaluations conducted in simulated Earth-Moon and Earth-Mars link scenarios demonstrate this algorithm's effectiveness in optimizing data transmission for interplanetary networks. Compared to existing methods, this approach demonstrates significant improvement, with matrix decoding failures reduced by at least 2/3.
\end{abstract}


\IEEEpeerreviewmaketitle

\section{Introduction}
\IEEEPARstart{T}{he} Delay/Disruption-Tolerant Networking (DTN)\\\cite{burleigh2003delay, fall2003delay} was proposed to address the challenges of interplanetary networks characterized by extremely long propagation delays and frequent link interruptions~\cite{akyildiz2003interplanetary}.
Within the DTN architecture, the Licklider Transmission Protocol (LTP)~\cite{rfc5326ltp, ccsds2015ltp} serves as a convergence layer protocol that aggregates upper-layer data into segments and provides reliable data delivery service using Automatic Repeat reQuest (ARQ)~\cite{lin1984automatic}.
Due to the significant propagation delays in deep space links\textemdash ranging from seconds in Earth-Moon scenarios to minutes or even hours in Earth-Mars scenarios~\cite{abdelsadek2022future}\textemdash each retransmission cycle incurs substantial time costs. 
Consequently, researchers have explored integrating packet-level forward erasure correction (PL-FEC) into LTP to enhance transmission performance further~\cite{ccsds2014fec}.

In~\cite{shi2017integration}, a novel protocol called RS-LTP is proposed, integrating Reed-Solomon (RS) codes~\cite{rfc6865RScodes} into LTP as a ``local data link layer" within the DTN protocol stack. Concurrently, The Erasure Coding Link Service Adapter (ECLSA)~\cite{apollonio2014erasure} was developed and later enhanced to ECLSAv2~\cite{alessi2019eclsa}. ECLSA is an intermediate-layer protocol beneath LTP, providing transparent protection for all upper-layer LTP segments using Low-Density Parity-Check (LDPC) codes~\cite{rfc6816ldpc}. These studies, implemented using the Interplanetary Overlay Network (ION)~\cite{burleigh2007ion}\textemdash an open-source DTN architecture software developed by NASA's Jet Propulsion Laboratory (JPL)\textemdash collectively demonstrate that integrating PL-FEC with LTP significantly enhances data transmission reliability and efficiency in challenging deep space communication environments, particularly under high bit error rates.

At the link layer, packet loss rates in satellite communications exhibit fluctuating patterns due to various environmental factors. In Ka-band RF communications, atmospheric effects, particularly rain attenuation, can cause deep fades lasting several minutes~\cite{igwe2019evaluation}. For optical satellite communications, atmospheric turbulence, mispointing errors, and cloud coverage lead to rapid fluctuations in received signal strength~\cite{kaushal2016optical}. Additionally, space weather phenomena such as solar winds and cosmic radiation introduce intermittent interference patterns in both RF and optical links. When these physical layer impairments exceed the correction capability of channel coding, bit errors manifest as burst packet losses at the link layer. This time-varying packet loss behavior suggests that link layer erasure coding strategies should adapt to these changing channel conditions rather than assume a static loss pattern.

RS-LTP and ECLSA employ typical block codes, where redundancy symbols are encoded across disjoint blocks of source information symbols. Retransmissions are requested only when the number of lost information symbols exceeds the recovery capability of the block. 
To avoid insufficient coding redundancy in time-varying channels, RS-LTP conservatively presets a low fixed code rate (typically 1/2) before transmission. However, this approach causes overprotection when channel conditions exceed worst-case assumptions, preventing optimal performance across varying conditions.
ECLSAv2 addresses this by dynamically adjusting code rates based on packet loss rates estimated from receiver feedback. However, performance issues persist in rapidly changing deep space links. Channel state changes cannot be detected until feedback arrives, potentially causing matrix decoding failures or redundancy waste during this period.
This issue becomes more severe in high-propagation-delay scenarios, where more data matrices are transmitted with mismatched parameters, significantly impacting system performance.

This study explores the application of reinforcement learning (RL) to help FEC-LTP overcome this performance bottleneck. RL has been extensively adopted in satellite communication fields such as adaptive coding modulation, congestion control, and routing algorithms in recent years. For instance, NASA marked a significant milestone by implementing machine learning on the International Space Station's SCaN testbed, demonstrating its effectiveness in space communications~\cite{ferreira2016multi,ferreira2017multi,ferreira2018multiobjective,ferreira2019reinforcement}. Although research on applying RL to PL-FEC has been relatively rare, it has recently attracted growing interest within the academic community. In~\cite{zhang2021learning}, a classical RL algorithm was effectively applied to drive streaming FEC coded transmissions, achieving high in-order goodput with low delay in a simulated geostationary earth orbit satellite link with random loss rates. In~\cite{latzko2023analysing}, a deep RL algorithm based on the deep Q-network (DQN) was employed to optimize random linear network coding selection strategies, improving communication performance across various tasks.

This paper focuses on the code rate control algorithm in FEC-LTP. To our knowledge, this study is the first to propose using RL to control PL-FEC for deep space communications. The following contributions are made in this paper.
\begin{enumerate}
    \item Based on the characteristics of interplanetary communication, we have designed and implemented an RL-based Adaptive PL-FEC algorithm, and applied it to the existing FEC-LTP implementation. In this algorithm, several new approaches including "Single-step Reinforcement Learning" and "State-action Buffer" are proposed to enhance RL learning and update strategies, making the RL scheme more suitable for the deep space communication scenario. Notably, these designs are not only applicable to FEC-LTP but also provide valuable reference for applying RL in deep space communications.
    \item In the Earth-Moon and Earth-Mars scenarios, we simulated time-varying links under realistic or extremely adverse conditions using two link packet loss models, followed by extensive evaluations of all schemes. Results indicate that the RL scheme learned the underlying mathematical distribution of link variations during training. When applied to actual transmissions, the RL agent assisted FEC-LTP in selecting more appropriate coding rates in most cases, significantly reducing decoding failures and thereby improving overall performance.
\end{enumerate}

\section{LTP and PL-FEC Integration for Enhanced Reliability}
This chapter provides readers with a brief introduction to LTP within the DTN architecture. It explains how PL-FEC is incorporated into LTP to enhance its transmission performance over long-delay and lossy links.

\subsection{DTN Architecture Overview}
\begin{figure}[!ht]
    \centering
    \includegraphics[width=8.5cm]{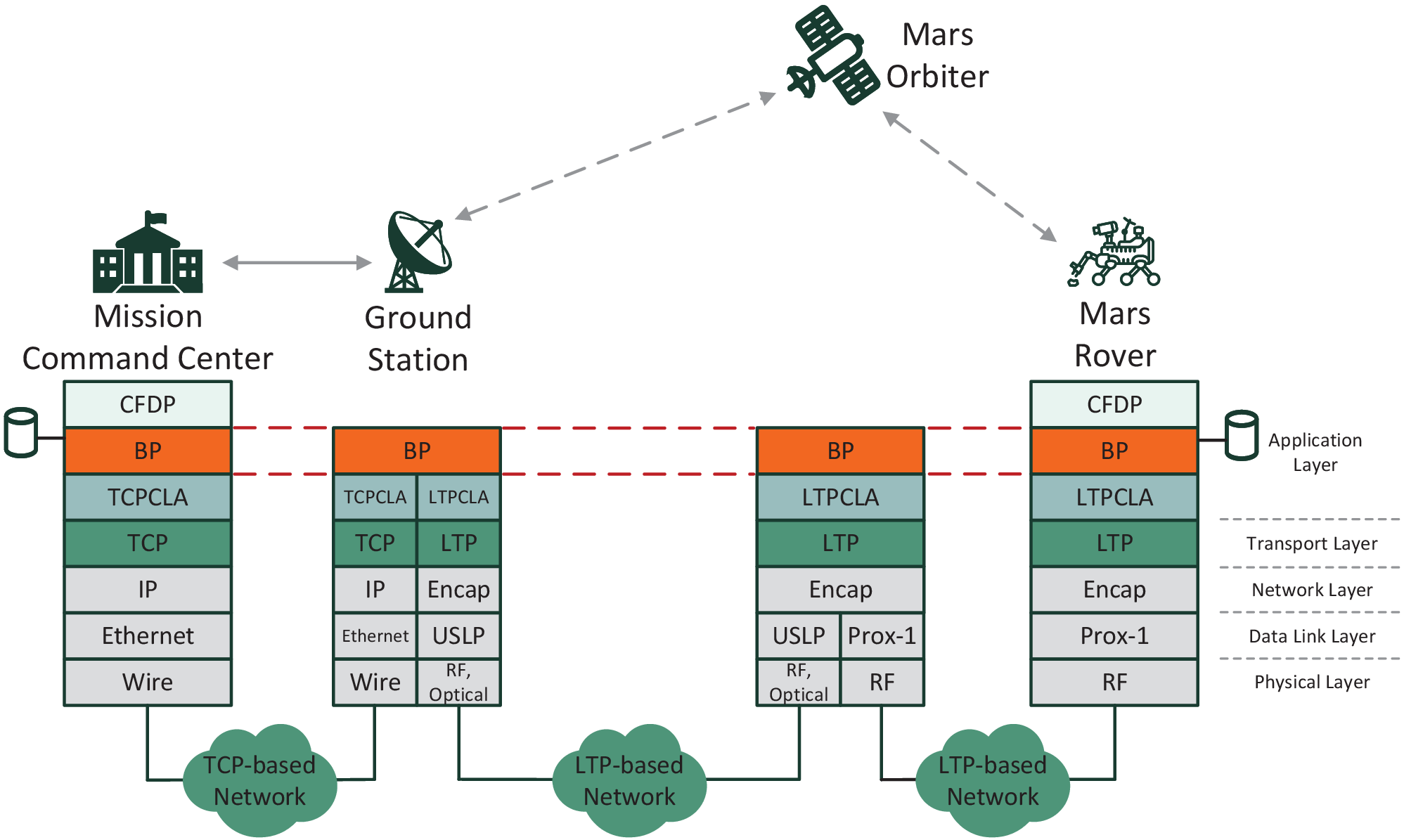}
    \caption{A protocol stack architecture example for a DTN-based Mars exploration mission.} 
    \label{DTN_stacks}
    \vspace{-5pt}
\end{figure}

Fig.~\ref{DTN_stacks} illustrates an example of DTN protocol stacks for Mars exploration missions. As shown, DTN architecture establishes a message-oriented, end-to-end overlay network by introducing a bundle layer between application and transport layer protocols. The corresponding bundle protocol (BP)~\cite{rfc5050bp, ccsds2015bp} is responsible for moving ``bundles" (packets at the homonymous layer) hop-by-hop between DTN nodes. By utilizing store-and-forward services provided by BP, bundles can be stored locally when the link between each DTN node and the next-hop node is interrupted and subsequently forwarded once the link is re-established. To a certain extent, this mechanism addresses the data transmission challenges caused by frequent link disruptions in deep space networks.

At each hop, nodes can deploy different underlying transport layer protocols to adapt to various transmission environments and provide the corresponding ``Convergence Layer Adapters" (CLAs) to upper BP for transmitting bundles. In the Earth-Mars communication scenario illustrated in Fig.~\ref{DTN_stacks}, two different CLAs are used by BP to interact with different network protocol stacks. TCP-based CLA (simply TCPCL)~\cite{rfc9174tcpcl} is used in the TCP-based network between the mission command center and the ground station. LTP-based CLA (simply LTPCL)~\cite{burleigh2013ltpcl} is used in the LTP-based network for deep space links (i.e., station--orbiter, orbiter--rover).

\subsection{Licklider Transmission Protocol}
LTP is a point-to-point transport protocol specifically designed to address the challenges of interplanetary communication~\cite{rfc5326ltp, ccsds2015ltp}. Since its introduction, LTP has attracted significant attention in space communications. NASA-JPL and many other academic institutions have conducted extensive research on LTP, focusing on performance evaluation~\cite{wang2024study, yang2023study, yang2019analysis, lent2018analysis}, parameter optimization~\cite{lent2022learning, shi2020study}, and mechanism improvements~\cite{alessi2019eclsa, lent2024evaluating, bisacchi2022multicolor, alessi2019design, shi2017integration}. These characteristics make LTP well-suited for deployment on long-delay interplanetary links.

\begin{figure}[!t]
    \centering
    \includegraphics[width=8.5cm]{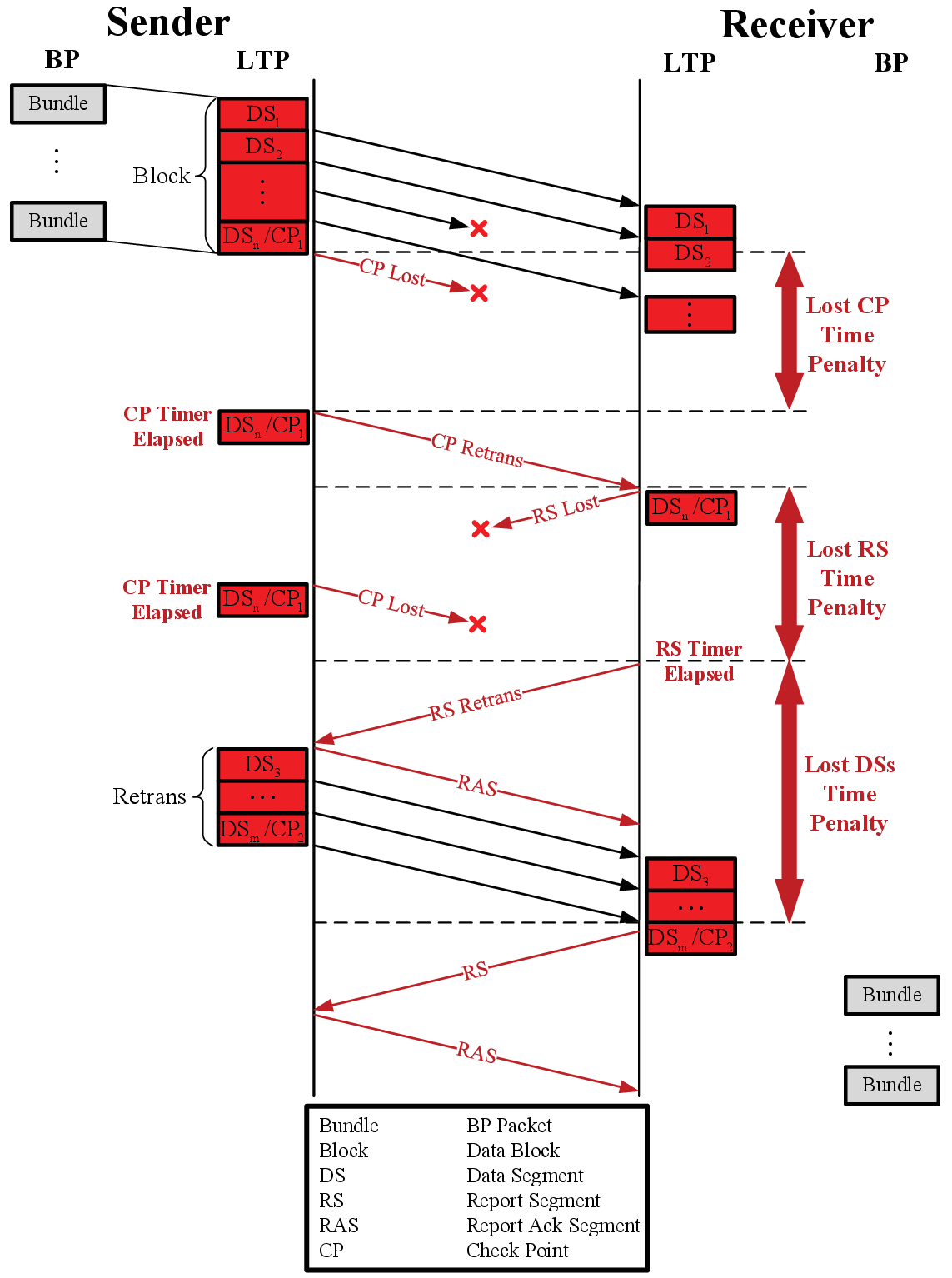}
    \caption{A data transmission scenario of LTP with BP illustrating the effect of segment loss on block transmission over lossy links.} \label{ltp_with_arq}
    \vspace{-10pt}
\end{figure}

LTP aggregates bundles from BP into data blocks and transmits them as segments in sessions. As shown in Fig.~\ref{ltp_with_arq}, LTP implements reliable delivery through an ARQ mechanism. The sender marks certain segments as checkpoints (CPs), which require receiver acknowledgment via Report Segments (RSs). Unacknowledged segments are retransmitted after timeout. This process continues until all segments are confirmed~\cite{rfc5326ltp}. Combined with LTP’s aggregation strategy, this approach reduces reverse acknowledgments, making LTP suitable for asymmetric links.

On long-delay, high-loss links, LTP's ARQ mechanism can introduce significant delays. Due to the transmission time of a data block is negligible compared to propagation delay, any segment loss triggers a retransmission cycle that adds one RTT delay penalty, as shown in Fig.~\ref{ltp_with_arq}. The situation worsens when control signals (CPs or RSs) are lost, causing session stalls until retransmission succeeds.

These challenges have prompted researchers to pursue improvements to the LTP mechanism. An advanced version has been developed that proactively replicates and transmits multiple copies of signaling segments~\cite{alessi2019design}, explicitly targeting the problem of signaling segment loss. Further approaches involve applying PL-FEC directly at the lower layer of LTP to provide coding protection for LTP segments~\cite{ccsds2014fec, alessi2019eclsa, shi2017integration}, thereby reducing the number of transmission cycles by utilizing additional bandwidth resources.


\subsection{PL-FEC Protection for LTP}
As mentioned in the introduction of this paper, ECLSA~\cite{alessi2019eclsa} has already implemented this idea and achieved encouraging test results. Unlike a simple Link Service Adapter that merely interfaces between LTP and lower-layer protocols like UDP, ECLSA functions as an intermediate-layer protocol providing transparent and equal FEC protection for all LTP segment types during transmission.

\begin{figure}[!t]
    \centering
    \includegraphics[width=8.5cm]{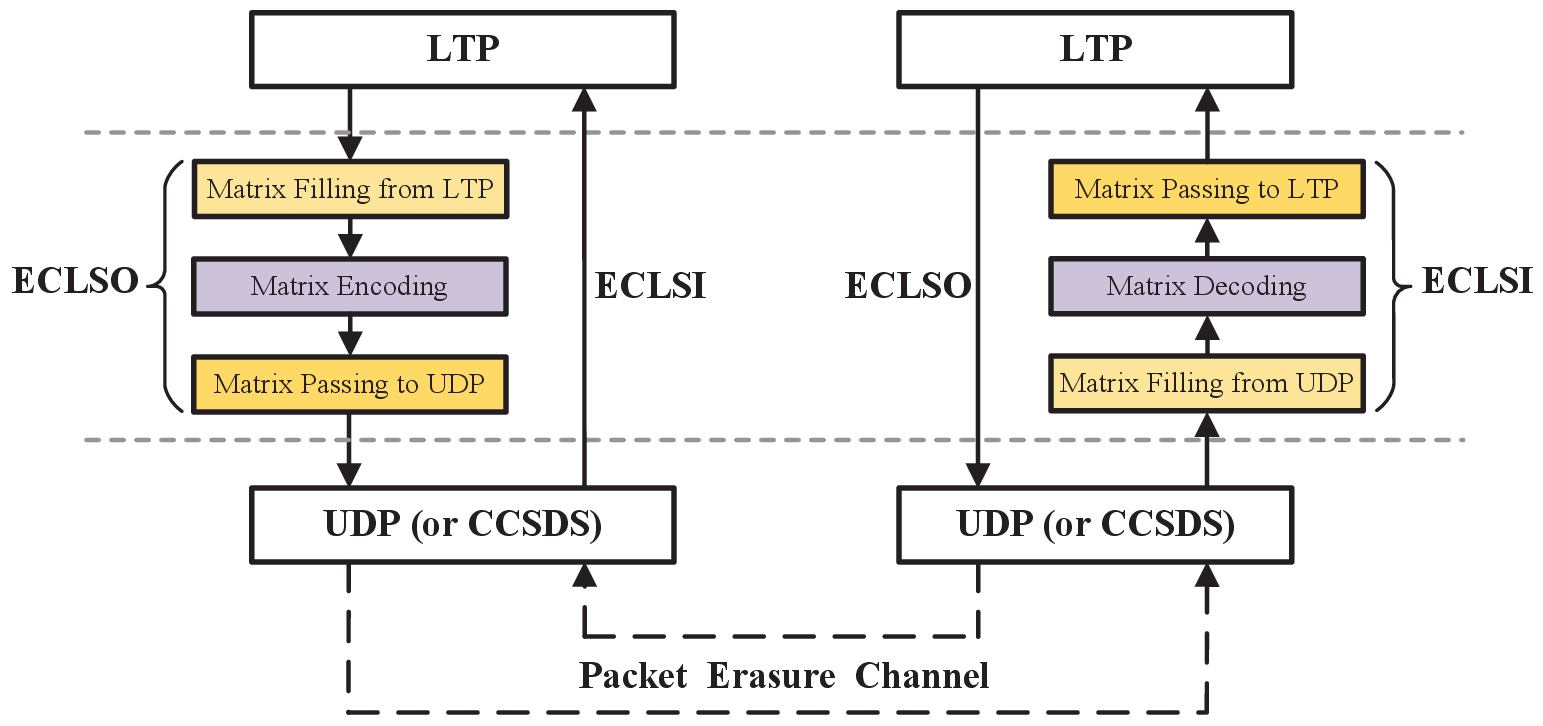}
    \caption{ECLSA implementation: ECLSO (left) and ECLSI (right) threads.}
    \label{eclsa_implementation}
    \vspace{-15pt}
\end{figure}

\begin{figure}[!t]
    \centering
    \includegraphics[width=8.5cm]{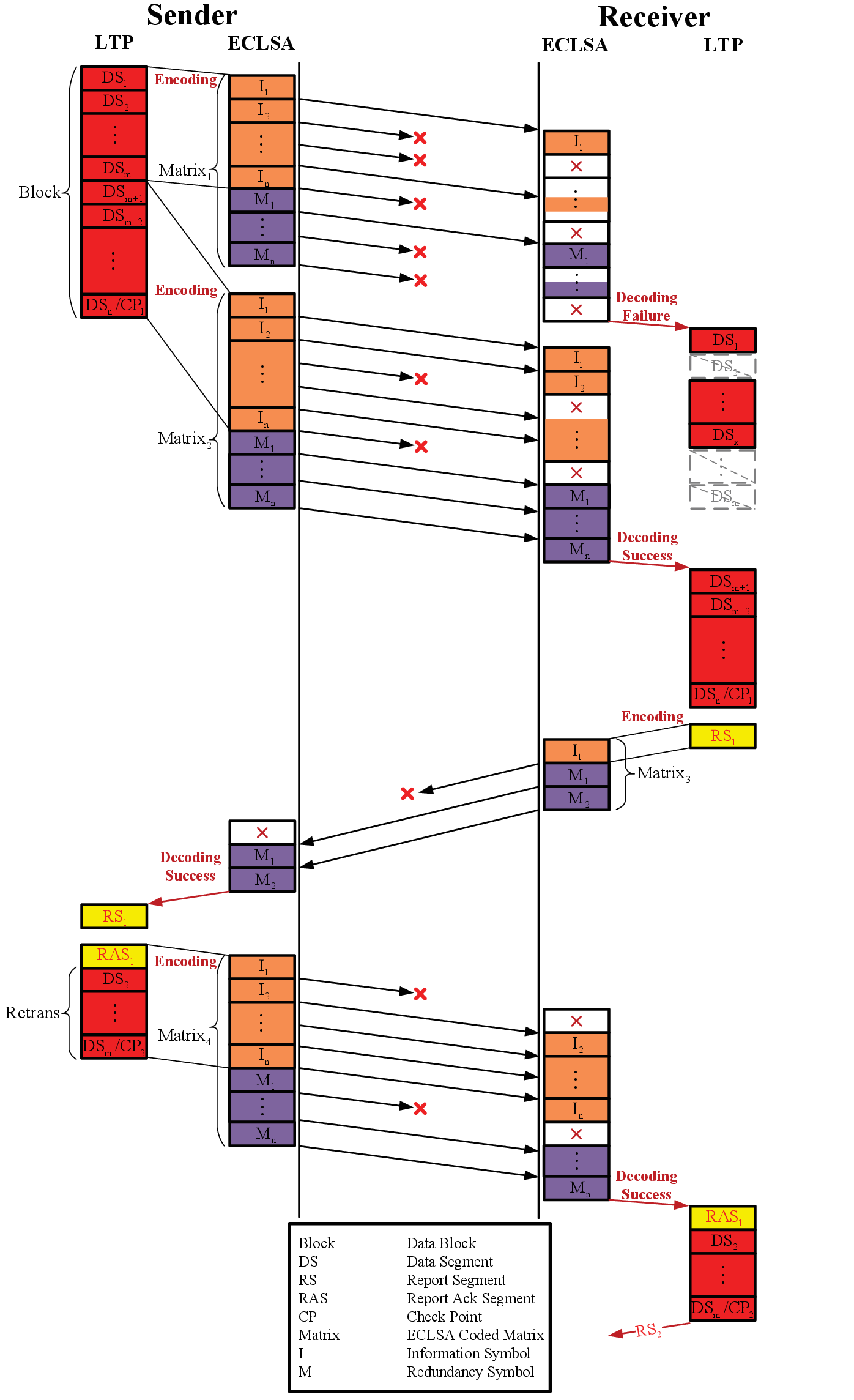}
    \caption{A scenario for LTP data block encoding transmission Using ECLSA to overcome link loss.}
    \label{ltp_with_eclsa}
    \vspace{-15pt}
\end{figure}

As shown in Fig.~\ref{eclsa_implementation}, the open-source ECLSA implementation in ION includes two processes: ECLSO and ECLSI, corresponding to the output and input channels~\cite[with changes]{alessi2020hsltp}. They perform complementary operations: the former encodes data segments received from LTP and passes them to the lower-layer protocol, while the latter decodes data received from the lower layer and passes it to LTP. Fig.~\ref{ltp_with_eclsa} presents an example of ECLSA provides coding protection under LTP to overcome link losses for a data block. We will use Fig.~\ref{eclsa_implementation} and~\ref{ltp_with_eclsa} to explain ECLSA's operation.

ECLSA leverages LDPC-based erasure coding with a coding matrix generating the $(N,K)$ codewords. On the sending side, LTP segments are inserted as information symbols into rows of the coding matrix. ECLSO is segment-oriented and does not consider LTP block boundaries, thus the data segments of a block may be distributed across multiple matrices. When a matrix is filled with $K$ segments or the aggregation timer expires, ECLSO generates redundancy symbols to form an $(N,K)$ codeword, adds headers to each symbol, and passes them to UDP.

On the receiving side, ECLSA reconstructs matrices from received symbols. If all K information symbols are received, decoding is skipped. Otherwise, ECLSA attempts to recover missing information symbols through decoding. When too many symbols are lost, the decoding may fail (as in $\text{Matrix}_{1}$), while successful recovery is possible with sufficient received symbols (as in $\text{Matrix}_{2}$). Regardless of decoding outcome, all received or recovered information symbols from each matrix are then delivered to the upper LTP layer.

For segments that ECLSA cannot recover, LTP's standard ARQ mechanism takes over. In Fig.~\ref{ltp_with_eclsa}, after receiving $\text{CP}_{1}$ from $\text{Matrix}_{2}$, the receiving-side LTP responds with $\text{RS}_{1}$. The local ECLSO encodes $\text{RS}_{1}$ into $\text{Matrix}_{3}$ for protection, which the receiving-side ECLSI successfully decodes. Upon receiving $\text{RS}_{1}$, the sending-side LTP responds with $\text{RAS}_{1}$ and retransmits the missing segments from $\text{Matrix}_{1}$. ECLSO encodes these into $\text{Matrix}_{4}$ and passes them to the receiving side. Successful decoding by the receiving-side ECLSI indicates the LTP block has been fully received.

\begin{figure*}[!t]
    \centering
    \includegraphics[width=.9\textwidth]{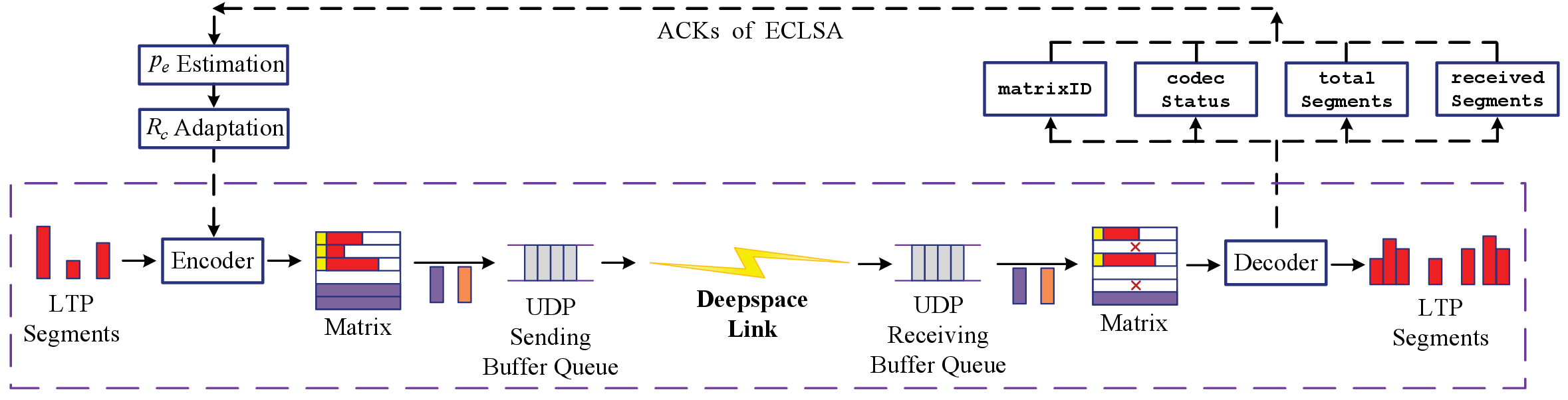}
    \caption{Data transmission process in feedback-adaptive \(R_c\) mode.}
    \label{feedback_rc_model}
    \vspace{-5pt}
\end{figure*}

As mentioned in the introduction, to reduce retransmission costs caused by decoding failures, ECLSA implements a delay-feedback-based adaptive coding algorithm, which requires enabling the ``feedback adaptive $R_c$" option in the ECLSO of sender.
Fig.~\ref{feedback_rc_model} illustrates the data transmission process with this algorithm enabled. After receiving and decoding each matrix, the receiver's ECLSI responds with feedback containing decoding status and reception statistics.

Upon receiving feedback, the sender's ECLSO adjusts the coding rate through the following steps:
\begin{enumerate}
    \item Calculate the symbol successful reception probability for this matrix:
    \begin{align}
        P_{\mathrm{s}} = \frac{\texttt{receivedSegments}}{\texttt{totalSegments}}
    \end{align}
    \item Interpret the \texttt{codecStatus}. If \textit{``Not Decoded"} or \textit{``Success"}, derive weight $w$ based on \texttt{totalSegments}, as follows:
    \begin{align}
        w = 
        \begin{cases}
            0.5, & \text{if } \texttt{total-} \\ 
            & \texttt{Segments} > 50 \\
            0.5 \cdot \frac{\texttt{totalSegments}}{50}, & \text{otherwise}
        \end{cases}
    \end{align}
    If \textit{``Failed"}, set $w$=1.
    \item Update estimated link packet loss rate \(p_e\): 
    \begin{align}
        p_{e,\mathrm{new}} = w \cdot (1-P_{\mathrm{s}}) + (1-w) \cdot p_{e,\mathrm{old}}
    \end{align}
    \item Update $R_c$: 
    \begin{align}
        R_c = 1 - p_e \cdot \mu
    \end{align}
    where $\mu$ is the expansion margin (default 1.15).
\end{enumerate}

Based on this algorithm, ECLSA can estimate $p_e$ in real-time at the sender and adjust $R_c$ accordingly. However, this strategy is ``passive adaptation." ECLSA cannot ascertain the latest $p_e$ until receiving feedback from matrices that have experienced the most recent link state. Consequently, $R_c$ adjustments lag behind link variations. Even after incorporating PL-FEC into LTP, performance bottlenecks persist when facing time-varying links. To address this, we propose a RL-based proactive adaptive coding algorithm, which will be detailed in the next chapter.

\section{Proposed Solutions} \label{proposed_solutions}
\begin{figure*}[!ht]
    \centering
    \includegraphics[width=.9\textwidth]{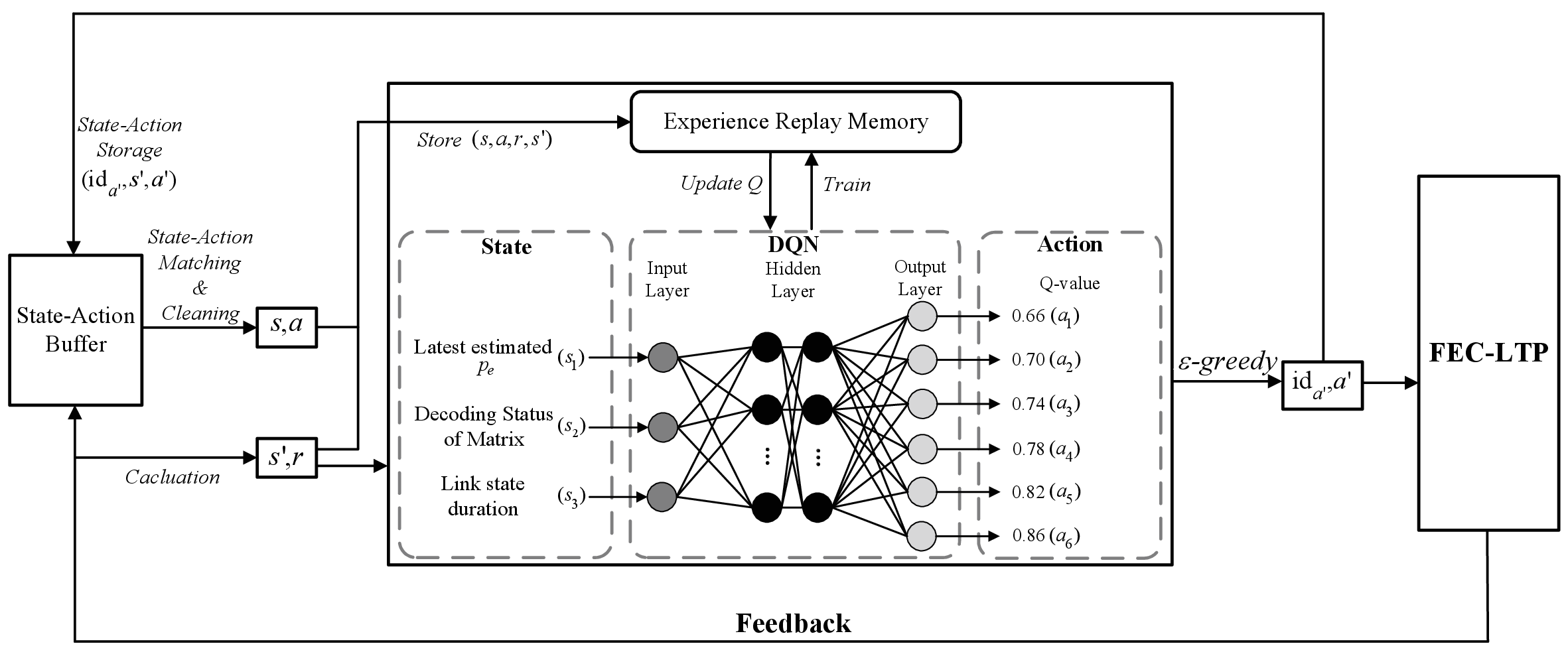}
    \caption{Internal Workflow of the RL Agent.}
    \label{system_model}
    \vspace{-5pt}
\end{figure*}

This chapter elaborates on the specific design and implementation of our proposed RL scheme, as well as its practical application within the existing FEC-LTP framework. Fig.~\ref{system_model} illustrates the internal workflow of the RL agent, which readers should refer to for a better understanding of the subsequent sections.

\subsection{Reinforcement Learning Model}
\begin{figure}[!ht]
    \centering
    \includegraphics[width=8.5cm]{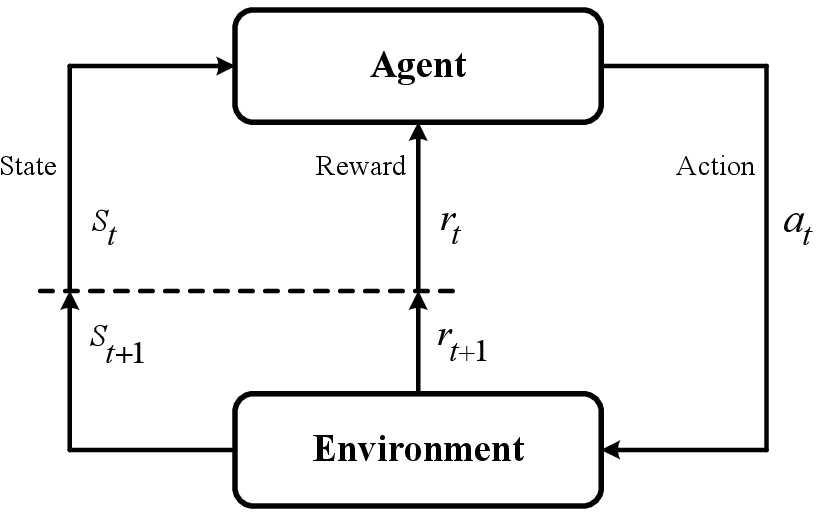}
    \caption{Network topology for performance evaluation.}
    \label{rl_example}
    \vspace{-5pt}
\end{figure}

Reinforcement Learning (RL) is a kind of distinctive machine learning paradigm that learns through interaction with an external environment and evaluating its performance based on the feedback~\cite{sutton1988reinforcement}. A RL task is commonly modeled using a Markov Decision Process (MDP)~\cite{busoniu2010reinforcement}, which can be briefly described as follows: An agent (i.e., the model in RL) operates within an environment, E. The state space, S, comprises states \(s \in S\) that represent the agent's perception of the environment. The action space, A, defines the possible actions available to the agent. When the agent takes an action \(a\) in the current state \(s\), the underlying state transition probability function determines the probability of transitioning to the next state \(s'\). The environment then provides feedback, \(r\), based on the reward function, indicating the ``reward" for the state transition resulting from action \(a\) in the state \(s\). This process continues iteratively, as illustrated in Fig.~\ref{rl_example}. As the interaction proceeds, the RL agent finally learns an optimal policy \(\pi\) for the given environment, which dictates the action \(a = \pi(s)\) to be executed in each state \(s\) to maximize cumulative rewards or reach a desired final state~\cite{ferreira2017multi}.

Using the environment model as prior knowledge for the RL agent is impractical for most of the practical applications due to the complexity of the environment. In this case, allowing the RL agent to operate without relying on an explicit environment model\textemdash namely model-free learning\textemdash becomes considerably important when applying RL methods. Temporal Difference (TD) learning is a kind of model-free learning methods employing a state-action value function \(Q(s, a)\) to represent the expected cumulative reward of taking action \(a\) in the state \(s\). The function is updated at each time step based on past experiences, according to the Q-function derived from the Bellman equation~\cite{sutton1988reinforcement, busoniu2010reinforcement}:
\begin{align}
Q_{k+1}(s_k, a_k) = Q_k(s_k, a_k) + \alpha [r + \gamma \max_a Q_k(s_{k+1}, a) \nonumber \\
    - Q_k(s_k, a_k)]
\end{align}
Here, $\alpha$ is the learning rate, $\gamma$ is the discount factor, and $r$ is the immediate reward. The policy $\Pi$ of Q-learning algorithm follows a greedy strategy:
\begin{align}
\pi(s) = \arg\max_{a} Q(s, a)
\end{align}
where, for every state \(s \in S\), an action \(a \in A\) with the maximal action-value is selected.

In this paper, we employ RL to control PL-FEC parameters in LTP, aiming to discover an optimal transmission policy that maximizes goodput, minimizes delivery delay, and reduces LTP retransmissions in a long-term dynamic communication environments.
These three metrics are strongly interrelated: goodput and delivery delay tend to be inversely proportional, and generally, a reduction in retransmissions can enhance the performance of the first two metrics. The RL agent aims to learn and master the underlying mathematical distribution of link variations through feedback, assisting FEC-LTP in real-time optimal code rate selection to improve these three performance metrics. Specifically, in this paper, the 3-component tuple of RL\textemdash action, state, and reward\textemdash are defined as follows:

\textbf{Action:}
The action given by the RL agent is the FEC code rate \(R_c\), which governs the PL-FEC encoding for the subsequent transmission period. To conform with the coding rate standards specified in the CCSDS orange book~\cite{ccsds2014fec}, the action space \(A\) is bounded between (2/3, 8/9), comprising six discrete values ranging from 0.66 to 0.86 with increments of 4\%, i.e., \(A = \{0.66, 0.70, 0.74, 0.78, 0.82, 0.86\}\). Finer code rate control is deemed unnecessary, considering the 2-3\% estimation error in packet loss rate \(p_e\) on the experimental link.

\textbf{State:}
The state sent to the RL agent in this paper has the physical meaning of ``the true state information about the communication link at the current time." The state \(s = \{s_1, s_2, s_3\}\) consists of three dimensions:
\begin{enumerate}
    \item \(s_1\): The latest estimated \(p_e\), a continuous value accurate to three decimal places. This value is obtained through ACK information transmitted from the FEC-LTP to the RL agent, as detailed in Section~\ref{subsec:implementation}.
    \item \(s_2\): The decoding status of the latest ACKed coding matrix, with values of 0 and 1, representing decoding failure and success, respectively.
    \item \(s_3\): The ``link state duration", signifying the time duration that the link has maintained the current \(p_e\) (i.e., the value of \(s_1\)). When feedback indicating the current link state arrives, the RL agent records the reception time as \(t_1\). When the next feedback arrives, it records the time as \(t_2\) and calculates the difference between the \(p_e\) estimates carried by these two consecutive feedbacks. If this difference exceeds a preset threshold, the RL agent determines that the link state has changed, indicating a significant variation in link quality. The duration of the new link state is initialized as:
\begin{align}
    \tau = t_2 - t_1
\end{align}
Otherwise, the link state is considered unchanged, and the duration is updated as:
\begin{align}
    \tau = \tau + t_2 - t_1
\end{align}
Note that at the RL agent's initialization, before receiving the first feedback, \(t_1\) is set as the output time of the first action.
\end{enumerate}

\textbf{Reward:}
Designing the reward function for deep-space communication requires careful consideration of practical task metrics and the agent's desired outcomes. This study focuses on optimizing goodput and file delivery time, primarily influenced by retransmission, coding redundancy, and receiver decoding overhead. Among these, decoding overhead, largely induced by link packet loss, is minimally affected by FEC code rates. In deep-space communication, throughput reduction from retransmissions significantly outweighs that from coding redundancy due to significant propagation delays. To address these challenges, we introduce \(\Delta\), a key parameter in our reward function:
\begin{align}
    \Delta = (1-p_e) - R_c
    \label{delta_function}
\end{align}
which measures the difference between the theoretically optimal code rate \(1-p_e\) and the chosen action code rate \(R_c\), based on real-time link condition feedback.

Building upon this, our reward function is designed as follows:
\begin{align}
    r =
    \begin{cases}
        \left(\frac{\Delta - 0.2}{0.2}\right)^2, & \text{if } \Delta > 0 \\
        -\left(\frac{\Delta}{0.2}\right)^2, & \text{if } \Delta < 0 \\
        -\left(\frac{\Delta}{0.2}\right)^2 - 1, & \text{if } \Delta < 0 \text{ and decoding fails}
    \end{cases}
    \label{reward_function}
\end{align}
This function reflects the ``accuracy" of the current action relative to the optimal code rate, encouraging actions approaching the theoretical optimum while penalizing significant deviations. To account for decoding failures' critical impact, an additional fixed penalty of 1 is imposed when receiver decoding fails.

Besides the 3-component tuple design, balancing exploration and exploitation is crucial for the RL algorithm. This study employs the \(\varepsilon\)-greedy algorithm described in \cite{sutton1988reinforcement}, where \(\varepsilon \in (0,1)\) represents the probability of the RL agent selecting a uniformly-distributed random action to explore the environment, balanced against a \(1-\varepsilon\) probability of exploiting acquired experience. Initially set to 1, \(\varepsilon\) decays by \(2 \times 10^{-4}\) per epoch during training until reaching a minimum of 0.01, transitioning from a purely random strategy to a more exploitative approach. This decay process is expressed as:
\begin{equation}
    \varepsilon = \max(0.01, 1 - 2 \times 10^{-4} \times \text{epoch})
\end{equation}
where \(\text{epoch}\) represents the current training epoch.

\subsection{DQN-Based Adaptive FEC Algorithm}
Over the past decade, the rapid development of Deep Learning has led to increased interest in Deep Reinforcement Learning (DRL), which combines Deep Neural Networks with RL techniques. This integration has become the leading approach in modern RL applications. In 2013, DeepMind pioneered the integration of NNs with RL, achieving remarkable results on Atari games~\cite{mnih2013playing}. This work culminated in publishing the first DRL algorithm, Deep Q-Network (DQN), in 2015~\cite{mnih2015human}. The DQN algorithm replaces the traditional Q-Learning's Q-Table with a NN. This approach computes Q-values for state-action pairs in real time via the NN, eliminating the need for explicit storage. In DQN, the current state \(s\) is fed into the NN as input and, after forward propagation, is mapped to outputs representing the Q-values of all actions in the state \(s\). Action selection follows the same greedy policy as in Q-learning. After executing the action and receiving environment feedback, DQN calculates the loss as the difference between \(Q_{\text{Target}}\) and \(Q_{\text{eval}}\) for backpropagation training. Here, \(Q_{\text{eval}}\) represents the current network's value estimate for all actions \(a\) in state \(s\), i.e., \(Q(s, a)\), obtained through forward propagation with \(s\) as input. \(Q_{\text{Target}}\) represents a more accurate value estimate based on current feedback, computed by:
\begin{align}
    r + \gamma \max_{a} Q_k(s_{k+1}, a)
\end{align}
where \(Q_k(s_{k+1}, a)\) is also obtained through forward propagation of the network. DQN iterates these processes via continuous interactions with the environment until convergence. It also introduces two key innovations, Experience Replay, and the Target Network, to enhance the convergence speed and the whole algorithm's stability.

The proposed PL-FEC control algorithm in this paper is based on the DQN framework. However, before delving into the algorithm's details, we must modify the above Q-function. In the Q-Learning algorithm, introducing the discount factor \(\gamma\) effectively represents the calculation of cumulative rewards, i.e., the rewards expected in the future by making more actions after taking a specific action at present. However, considering deep space communication tasks, each transmission round is physically independent, meaning that the performance of each transmission does not affect or is not affected by the performance of previous or future transmissions. In this context, the RL agent should be more interested in the immediate reward \(r\) and strive to achieve the best performance in the next transmission round based on existing experience and knowledge. The discount factor \(\gamma\) thus loses its practical significance in this case, and the Q-function should be revised as:
\begin{align}
    Q_{k+1}(s_k, a_k) = Q_k(s_k, a_k) + \alpha [r - Q_k(s_k, a_k)]
\end{align}

This concept is referred to as ``one-step reinforcement learning," which focuses solely on maximizing the immediate reward without considering the interdependence of actions in RL. Consequently, for DQN, the corresponding loss function should also be modified to:
\begin{align}
    r - Q_{\text{eval}}
\end{align}

where \(r\) is \(Q_{\text{Target}}\). A comprehensible description of this expression is that the network's estimation of state-action value should closely reflect the immediate reward obtained. In this context, since \(Q_{\text{Target}} = r\) is informed by the environment feedback, it is unnecessary to employ the target network mentioned in \cite{mnih2015human} for forward propagation, thereby simplifying the algorithm to require only a single NN.

Additionally, this paper incorporates the Dueling-DQN approach \cite{wang2016dueling}, where two branches, \(V\) and \(A\), are extended from the output of the fully connected hidden layer. \(V\) represents the value of the current state, while \(A\) represents the ``advantage" of each predicted action. The Q-values of each action are ultimately calculated by combining these two branches. The architecture of Dueling-DQN enables a certain degree of separation between the highly correlated parameters ``state" and ``action" in RL. During NN forward propagation, the state is no longer entirely dependent on actions but acquires its value prediction. In the interaction process, the agent learns not only the action values in different states but also the value of each state itself.

In summary, we employed an NN architecture consisting of a total of five layers: an input layer that has three neurons corresponding to the three dimensions of the state; two fully connected hidden layers each have 256 neurons, both using Relu as the transfer function; the output of the hidden layer branches into \(V\) and \(A\), consisting of one and six neurons respectively; the final output layer contains six neurons, corresponding to the possible six code rate values. According to the \(\varepsilon\)-greedy strategy, the RL agent selects the action with the maximum Q-value from the network output as the following action with a probability of \(1-\varepsilon\) or selects a random action with a probability of \(\varepsilon\). Every 100 actions are defined as one episode, and the RL algorithm's training convergence is determined by averaging the rewards over the most recent ten episodes. Once this average reward becomes stable over time, the training of the NN is considered complete.

\begin{algorithm}
\renewcommand{\thealgorithm}{}
\caption{DQN-based Adaptive FEC Control Algorithm} 
\begin{algorithmic}[0]
\State Initialize replay memory $\mathcal{D}$ to capacity $N$
\State Initialize state-action buffer $\mathcal{B}$ to capacity $P$
\State Initialize action-value function $Q$ with random weights $\theta$
\State Initialize exploration rate $\epsilon \gets 1$
\State Initialize ActionID $C \gets 0$
\For{episode $\gets 1$ to $M$}
    \State Initialize sequence $\mathbf{s_1} \gets \{x_1, x_2, x_3\}$
    \For{$t \gets 1$ to $T$}
        \State With probability $\epsilon$ select a random action $a_t$
        \State otherwise select $a_t = \arg\max_a Q(s_t, a; \theta)$
        \State Execute action $a_t$ and store $(\mathcal{C}, s_t, a_t)$ in $\mathcal{B}$
        \State Observe new state $\bm{s_{t+1}}$ and match it with 
        \State $(\mathcal{C}, s_p, a_p)$ in $\mathcal{B}$
        \State Calculate action accuracy $\Delta$ using equation (10)
        \State Calculate reward $r_t$ using equation (11)
        \State Store transition $(s_p, a_p, r_t, s_{t+1})$ in $\mathcal{D}$
        \For{$q \gets 0$ to $p-1$}
            \State Remove $(C, s_q, a_q)$ from $\mathcal{B}$
        \EndFor
        \State \textbf{end for}
        \State $C \gets C + 1$
        \State Sample random minibatch of transitions from $\mathcal{D}$
        \State Set $y_j \gets r_j$
        \State Perform a gradient descent step with Adam  
        \State optimizer on $(y_j - Q(s_h, a_h; \theta))^2$ 
        \State Update the network parameters $\theta$
    \EndFor
    \State \textbf{end for}
\EndFor
\State \textbf{end for}
\end{algorithmic}
\end{algorithm}

\subsection{State-Action Buffer for Delayed Feedback} \label{subsec:state-action-buffer}
\begin{figure}[!t]
    \centering
    \includegraphics[width=8.5cm]{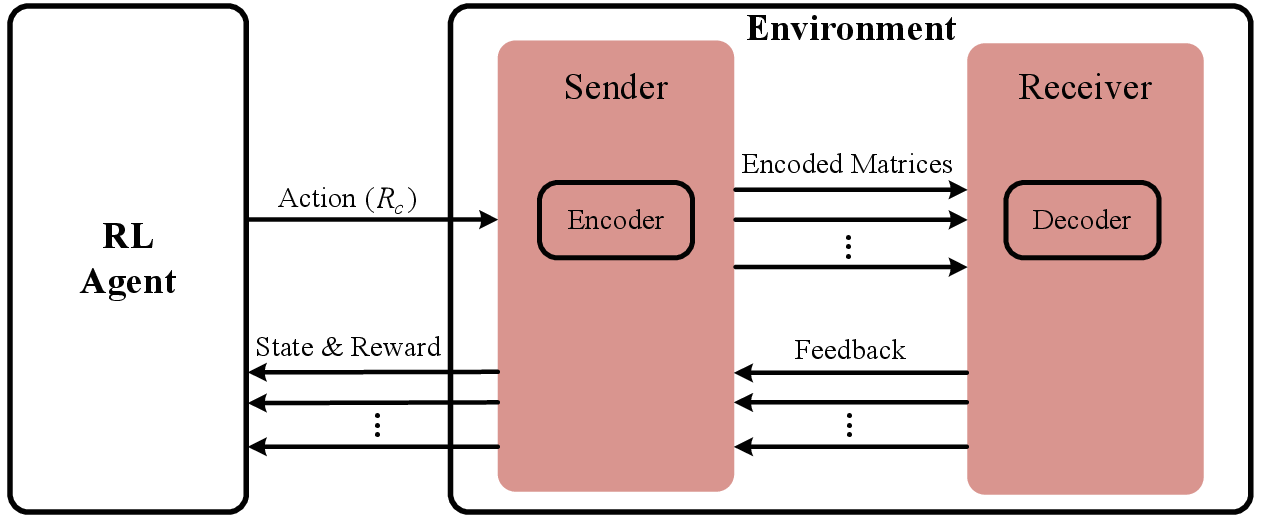}
    \caption{Single action from RL agent corresponding to multiple state-reward pairs.}
    \label{state}
    \vspace{-10pt}
\end{figure}

In traditional RL scenarios, an agent typically waits for environmental feedback (state and reward) after each action before selecting the following action. However, the deep space communication scenario using LTP deviates from this model owing to its non-wait-stop mechanism. Due to the scarcity of deep space communication resources, LTP typically initiates multiple parallel sessions for simultaneous transmission. The FEC layer beneath LTP needs to encode and transmit multiple matrices for each transmission round. The encoding and passing of a subsequent matrix do not need to wait for feedback from the previous matrix. Consequently, due to the long delays in communication links, a single action output from the RL agent guides the transmission of multiple matrices. The agent will receive multiple feedback corresponding to this single action, as illustrated in Fig.~\ref{state}.

One straightforward solution to address this ``Delayed Feedback" issue is that the RL agent only responds to and learns from the first feedback packet received for each action. However, this approach presents several challenges: training typically requires three or four feedback packets before outputting the next action, leading to reduced training efficiency; since a single action guides multiple transmissions, an inaccurate action incurs significant costs in decoding failures or bandwidth waste. Consequently, this approach is not suitable as a practical solution.

To ensure training efficiency while solving the problem of delayed action feedback, we propose a ``State-Action Buffer" design (from now on referred to as ``buffer") coupled with an ``Action ID" mechanism. This approach stores actions awaiting feedback, facilitating matching each action with its corresponding feedback packet. The operational logic comprises three steps:

\textbf{State-Action Storage:}
When the RL agent selects an action $a$ in state $s$, a triplet $(\mathrm{id}_a, s, a)$ is stored in the buffer. The $\mathrm{id}_a$ is a global identifier that initializes at 0 and increments by one with each action during a single run, indicating the sequential number of the action output by the RL agent. When transmitting information to FEC-LTP, the RL agent sends the action $a$ (i.e., $R_c$) and its corresponding identifier $\mathrm{id}_a$. The FEC-LTP sender records the $\mathrm{id}_a$ corresponding to the $R_c$ used for each encoded matrix. Upon receiving feedback for a matrix from the receiver, the FEC-LTP sender relays this feedback along with the corresponding $\mathrm{id}_a$ to the RL agent.

\textbf{State-Action Matching:}
Once receiving a feedback packet from FEC-LTP, the RL agent extracts the embedded $\mathrm{id}_a$. This identifier retrieves the corresponding state $s$ and action $a$ from the buffer, thus determining which state-action pair this feedback corresponds to.

\textbf{State-Action Cleaning:}
Due to the ``first-in-first-out" (FIFO) characteristic of ECLSA's operation, feedback packets for earlier transmitted matrices invariably arrive first. Consequently, the buffer clears all triplets $(\mathrm{id}_a, s, a)$ with $\mathrm{id}_a$ values smaller than that in the current received feedback packet, thereby freeing storage resources.

With this design, the RL agent doesn't wait for feedback on each action before selecting the following action. Instead, it stores the action in the buffer, awaiting feedback. Upon receiving a feedback packet, the RL agent locates its corresponding state-action pair in the buffer and then proceeds with updates and training. This approach significantly reduces training overhead as the RL agent selects an action and trains for every received feedback packet. As the number of feedback packets equals the number of received encoded matrices, and each feedback packet corresponds to a state and generates a corresponding action, a single action typically doesn't guide multiple matrix transmissions. This reduces performance loss caused by a single erroneous action \footnote{An exception occurs at the very start of transmission when no feedback has arrived, allowing the first action to guide multiple transmissions.}. Additionally, due to the ``State-Action Cleaning" step, the number of triplets simultaneously stored in the buffer is typically limited (related to FEC-LTP's processing delay and link RTT). Thus, introducing the buffer does not impose significant storage or search overhead on the system.

\subsection{Implementation} \label{subsec:implementation}
Given ECLSA's excellent performance in FEC-LTP implementation, this study chooses to implement the proposed RL scheme based on ECLSA. ECLSA was integrated into ION-DTN and deployed during the initial compilation and installation. ECLSA and ION-DTN are developed in C, while the RL model is built using Python. To address the information exchange between ECLSA and the RL agent, this study employs Unix domain sockets as a local inter-process communication mechanism for transmitting $R_c$ and feedback of matrices. Local Unix communication, bypassing the protocol stack, has extremely low latency (tested at approximately 1~ms), ensuring timely matrix encoding without significant delay in $R_c$ transmission.

To this end, we modified the source code of ECLSA, adding a ``RL adaptive $R_c$" option. When enabled, ECLSO creates a local Unix socket and launches a dedicated thread at a preconfigured communication address, waiting in blocking mode to receive action from the RL agent, which is then unconditionally adopted.

As described in Section~\ref{subsec:state-action-buffer}, the RL agent requires FEC-LTP to store $\mathrm{id}_a$ and transmit them along with matrix feedback. To meet this requirement, the modified ECLSA in RL mode implements a ``Matrix Information Buffer" indexed by matrix ID $\mathrm{id_{matrix}}$ on the ECLSO. This buffer manages the storage and retrieval of relevant information for each matrix through the following steps:
\begin{enumerate}
    \item After matrix encoding, it records:
    \begin{itemize}
        \item $\mathrm{id_{matrix}}$
        \item $\mathrm{id}_a$
        \item $I$, $K$, and $N$
    \end{itemize}
    
    \item After matrix passing, it records:
    \begin{itemize}
        \item Transmission time (ms)
    \end{itemize}
    
    \item Upon receiving the corresponding feedback, based on \texttt{matrixID}, it records:
    \begin{itemize}
        \item \texttt{TotalSegments}
        \item \texttt{ReceivedSegments}
    \end{itemize}
    
    \item ECLSO transmits the complete transmission information of this matrix to the RL agent via Unix socket.
    
    \item ECLSO clears information for all matrices\textemdash those that are awaiting feedback\textemdash with IDs smaller than the current matrix ID.
\end{enumerate}

This design enables the RL agent to calculate the state and reward corresponding to each action based on information provided by ECLSA, select the optimal action, and timely transmit it to ECLSA, thereby achieving adaptive $R_c$ adjustment.

After integrating the RL model with FEC-LTP, the next step is to apply it to real-world link environments. However, given the RL model's $\varepsilon$-greedy policy, we posit that the agent's actions essentially explore link characteristics until convergence. Considering the critical role of coding rates in FEC-LTP's data transmission efficiency over interplanetary links, we recommend a cautious implementation strategy. Extensive offline training of the RL model in actual link environments should precede practical application. The primary objective is for the RL agent to learn and master the underlying mathematical distribution of link variations, enabling it to assist FEC-LTP in real-time selection of optimal coding rates for efficient file transfer in subsequent real-world applications.

\section{Experimental Testbed and Configuration}
This chapter details the experimental testbed development for comparing the proposed RL method with traditional approaches in a simulated environment. It covers the simulation of interplanetary links, design of link packet loss rate \(p_e\) variation models, and configuration of the DTN protocol stack deployed using ION software.

\subsection{Simulating Interplanetary Links}
\begin{figure}[!ht]
    \centering
    \includegraphics[width=8.5cm]{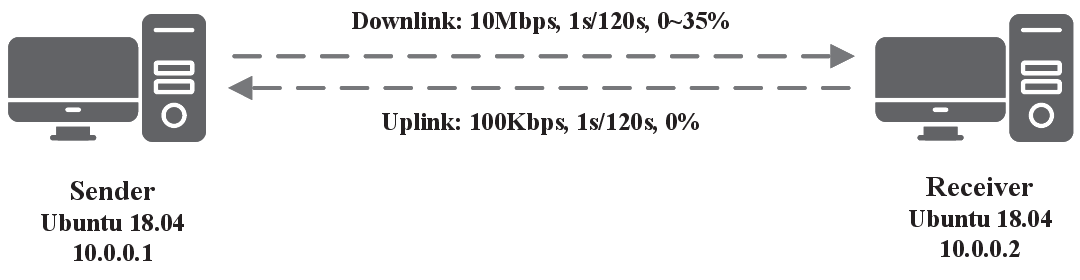}
    \caption{Network topology for performance evaluation.}
    \label{network_topology}
    \vspace{-10pt}
\end{figure}

Evaluation is conducted on a 2-node linear network as shown in Fig.~\ref{network_topology}. All nodes run Ubuntu 18.04 LTS (Linux kernel v5.4.0). Interplanetary links between nodes are simulated using Linux \texttt{tc-tbf} and \texttt{tc-netem} to configure each node's network interface.

Previous research~\cite{users2007revision} indicates that space communication asymmetry ratios range from a minimum of 50:1 to a maximum of 1000:1. To simulate a high channel asymmetry ratio of 100:1, downlink and uplink bandwidths are configured at 10Mbps and 100Kbps, respectively.

Current interplanetary exploration missions focus primarily on lunar and Martian targets. The one-way propagation delay between Earth and the Moon is approximately 1280 ms, while the Earth-Mars delay ranges from 3 to 22 minutes. To simulate these two scenarios, the one-way propagation delay between the nodes is set to 1 s or 120 s, representing lunar and Martian link conditions, respectively. While these values differ from actual delays, they do not compromise the experiment's validity. The primary goal is to verify performance differences among all schemes under varying delay magnitudes rather than precisely simulating specific distance communication links. This setup enables more transmission experiments within limited computational resources and time, enhancing the statistical significance of results.

\subsection{Link Packet Loss Models}
To simulate dynamic link quality, we designed two models controlling real-time variations in \(p_e\).

\subsubsection{Discrete Uniform Distribution}
$p_e$ follows a discrete uniform distribution, where each value is randomly selected with equal probability from a finite sample set $\mathcal{P}$. This probabilistic model effectively simulates unpredictable random variations in link quality across different discrete levels. According to the coding rate range (2/3, 8/9) specified in the CCSDS Orange Book~\cite{ccsds2014fec}, the set \(\mathcal{P}\) is defined as:
\begin{equation}
    \mathcal{P} = \{0\%, 5\%, 10\%, 15\%, 20\%, 25\%, 30\%, 35\%\}
\end{equation}

The time intervals for \(p_e\) changes are fixed. In our experiments, this model was employed only in the Earth-Moon scenario, with the interval set to 5 RTTs or 10 RTTs.

\subsubsection{Markov Chain}
\begin{figure}[!ht]
    \centering
    \includegraphics[width=8.5cm]{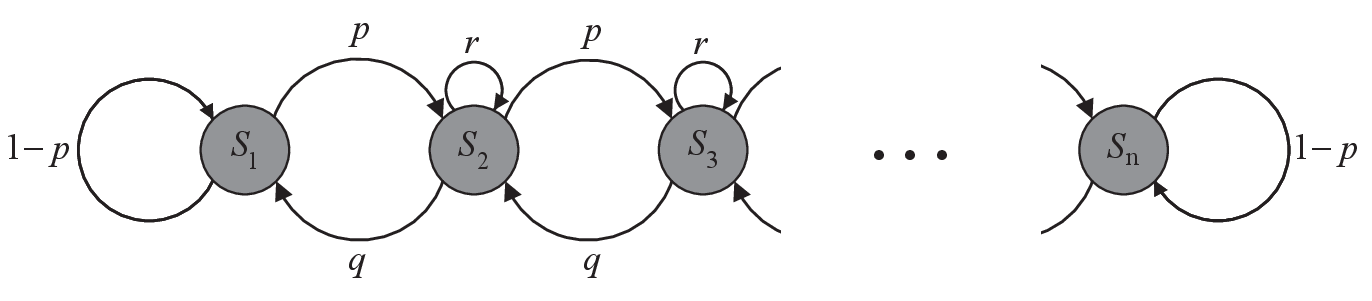}
    \caption{Multi-state Markov chain model.}
    \label{markov_chain}
    \vspace{-10pt}
\end{figure}

A Markov chain is a mathematical model describing a system's random state transitions over time, exhibiting ``memorylessness" where transitions depend only on the current state. Compared to the Gilbert-Elliott channel model (a two-state Markov chain), this model achieves more detailed characterization of channel variations through multiple states, making it widely used in satellite channel modeling~\cite{lopez2019finite,palmieri2022proposal,tropea2022comprehensive}. This study employs a discrete-time first-order multi-state birth-death Markov chain with holding states (Fig.~\ref{markov_chain}).

In this model, the system can only transition to adjacent states or remain unchanged, where \(p+q+r=1\). By setting \(p_e\) to follow this model, we simulate more realistic link quality variations, effectively capturing temporal correlations and gradual changes in network states. The state space is defined as:
\begin{equation}
    \mathcal{S} = \{5\%, 15\%, 20\%, 25\%, 30\%, 35\%\}
\end{equation}
Compared to the discrete uniform distribution model, we removed the 0\% and 10\% states. For the highest code rate 8/9, 0-15\% loss changes don't trigger code rate adjustments in all schemes. This Markov chain model prohibits random jumps, so we chose the 5\% state to represent the 0-15\% interval.

The state transition matrix $P$ is defined as:
\begin{equation}
P = \begin{bmatrix}
0.4 & 0.6 & 0 & 0 & 0 & 0 \\
0.4 & 0.2 & 0.4 & 0 & 0 & 0 \\
0 & 0.4 & 0.2 & 0.4 & 0 & 0 \\
0 & 0 & 0.4 & 0.2 & 0.4 & 0 \\
0 & 0 & 0 & 0.4 & 0.2 & 0.4 \\
0 & 0 & 0 & 0 & 0.6 & 0.4
\end{bmatrix}
\end{equation}

The time intervals for \(p_e\) state transition follow an exponential distribution with parameter \(\lambda\), exhibiting memorylessness consistent with Markov chains. For the expected value \(\frac{1}{\lambda}\), we set up 5 RTTs and 10 RTTs for the Earth-Moon scenario and only 5 RTTs for the Earth-Mars scenario.

File transfer experiments were conducted exclusively on the downlink. The uplink transmitted only ACKs, such as LTP signaling segments and ECLSA feedback. Not exceeding 200 bytes, these packets can be reliably transmitted via ECLSA's encoding and feedback replication. Consequently, the two models mentioned above were applied solely to the downlink. Dynamic \(p_e\) variations were simulated by a Python script on the sender node, automatically executing Linux \texttt{tc change} commands to modify network interface conditions.

\subsection{DTN Protocol Stack Based on ION}
The experimental protocol implementation of BP/LTP was adopted from the ION distribution v4.0.1 with ECLSA implementation. The underlying network and link layers employed IP and Ethernet, respectively. 
As this study primarily focuses on controlling the FEC layer beneath LTP, DTN protocol parameters were configured with fixed values across all experiments, as shown in Table~\ref{tab:protocol_parameters}.
\begin{table}[!ht]
\renewcommand{\arraystretch}{1.3}
\centering
\caption{Protocol Parameters Settings}
\label{tab:protocol_parameters}
\begin{tabularx}{0.48\textwidth}{
    >{\centering\arraybackslash}X
    | >{\centering\arraybackslash}X
    | >{\centering\arraybackslash}X
}
\hline
Protocols & Parameters & Values \\
\hline\hline
\multirow{4}{*}{LTP} & Block Size & 600\,000 B \\
 & Segment Size & 1024 B \\
 & \multirow{2}{*}[0.3ex]{\makecell{Max Export /\\Import Sessions}} & \multirow{2}{*}[0.3ex]{\makecell{5 (Earth-Moon),\\30 (Earth-Mars)}} \\
 & & \\
\hline
\multirow{3}{*}{ECLSA} & $K_\text{max}$ & 512 \\
 & $N_\text{max}$ & 768 \\
 & $T_\text{max}$ & 1026 B \\
\hline
\end{tabularx}
\small
\begin{tablenotes}
\item Note: LTP = Licklider Transmission Protocol, ECLSA = Erasure Coding Link Service Adaptor,
\item B = Bytes
\end{tablenotes}
\end{table}
\vspace{-10pt}

\section{Experimental Results and Analysis}
We conducted comprehensive performance comparison tests between ECLSA's feedback-adaptive $R_c$ algorithm and our proposed RL-adaptive $R_c$ algorithm by deploying each scheme on the experimental testbed and executing file transfer tasks of 50MB each. Our testing process comprised two distinct phases: a training phase for the RL scheme and a formal testing phase for both methods. During the training phase, exclusive to the RL scheme, we executed approximately 100 rounds of file transfer tasks in each scenario. This phase continued until the average reward per episode reached the convergence criteria outlined in Chapter~\ref{proposed_solutions}. The formal testing phase, applied to both methods, consisted of 100 rounds of file transfer tasks for each method in every scenario.

The link packet loss rate control script was executed before the start of the transmission experiments and remained active until after the completion of the final round. Consequently, each round of file transmission experienced different link variations.

Our objective is to enable the RL agent to learn the underlying mathematical distribution of link variations during training. This should allow it to perform excellently when placed in a new link following the same distribution. To verify this, we used different random number seeds in experiments, ensuring that the RL agent experienced distinct link variation processes during the training and formal testing phases. This approach allowed us to assess whether the RL agent truly learned the underlying distribution rather than merely memorizing specific loss patterns. During the formal testing phase, we set identical random number seeds across all schemes. This ensured that all approaches experienced exactly the same link variation processes during testing, guaranteeing a fair comparison.

To thoroughly evaluate the performance of both methods, we focused on three key metrics: goodput (Mbps), file delivery delay (seconds), and the number of decoding failures. After each round of file transmission, the first two metrics are directly calculated and provided by ION. 
The third metric is obtained by analyzing ECLSA's log file and counting the number of matrices that failed to decode. In the subsequent sections, we present a detailed analysis of the test results for each scenario, offering insights into the comparative performance of the two adaptive $R_c$ algorithms under various conditions.

\begin{figure*}[!t]
\begin{minipage}[t]{0.5\textwidth}
    \centering
    \begin{subfigure}[b]{.48\textwidth}
        \caption{\centering}
        \label{earth-moon-random-5RTT:goodput}
        \vspace{-2ex}
        \includegraphics[width=\textwidth]{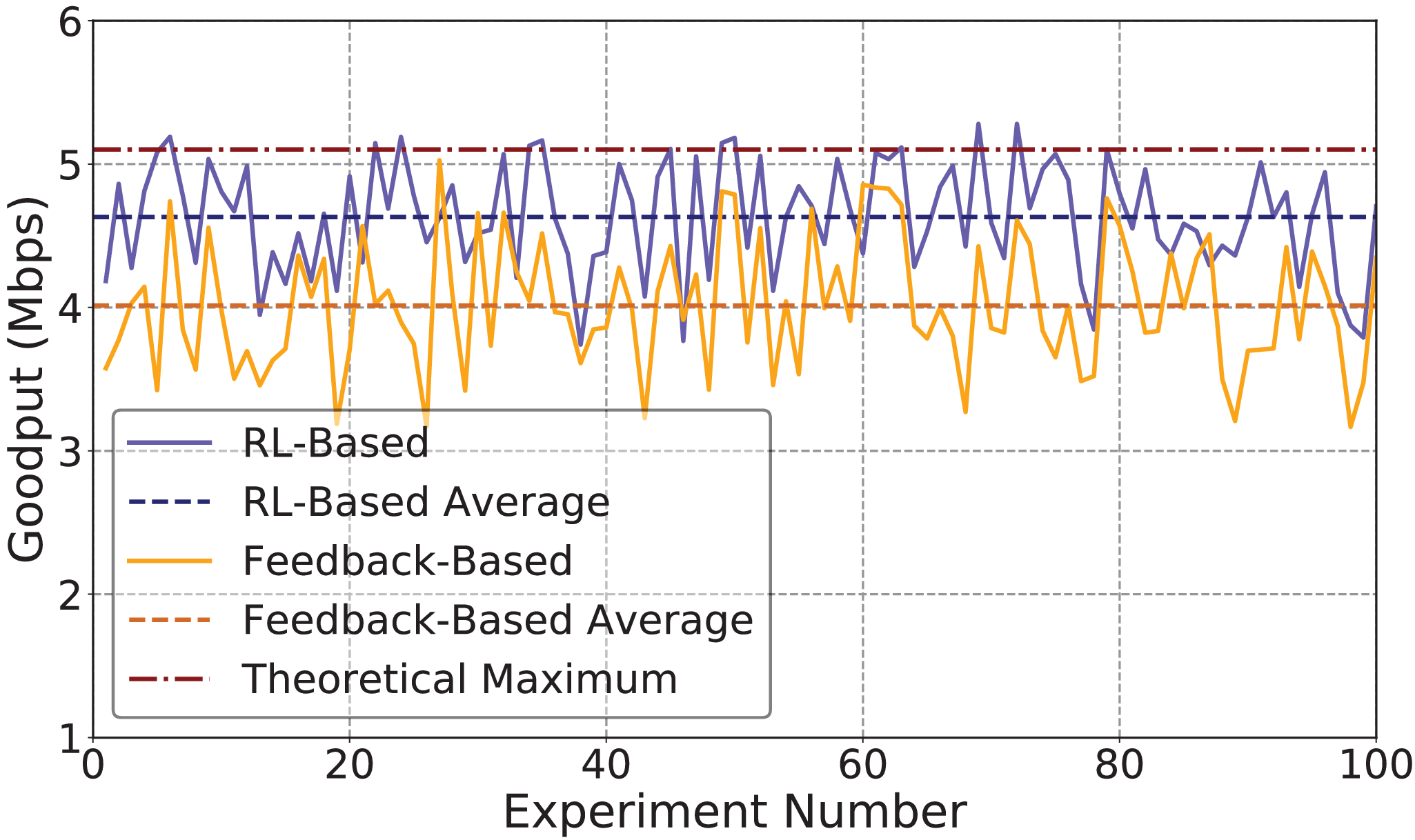}
    \end{subfigure}
    \hfill
    \begin{subfigure}[b]{.48\textwidth}
        \caption{\centering}
        \label{earth-moon-random-10RTT:goodput}
        \vspace{-2ex}
        \includegraphics[width=\textwidth]{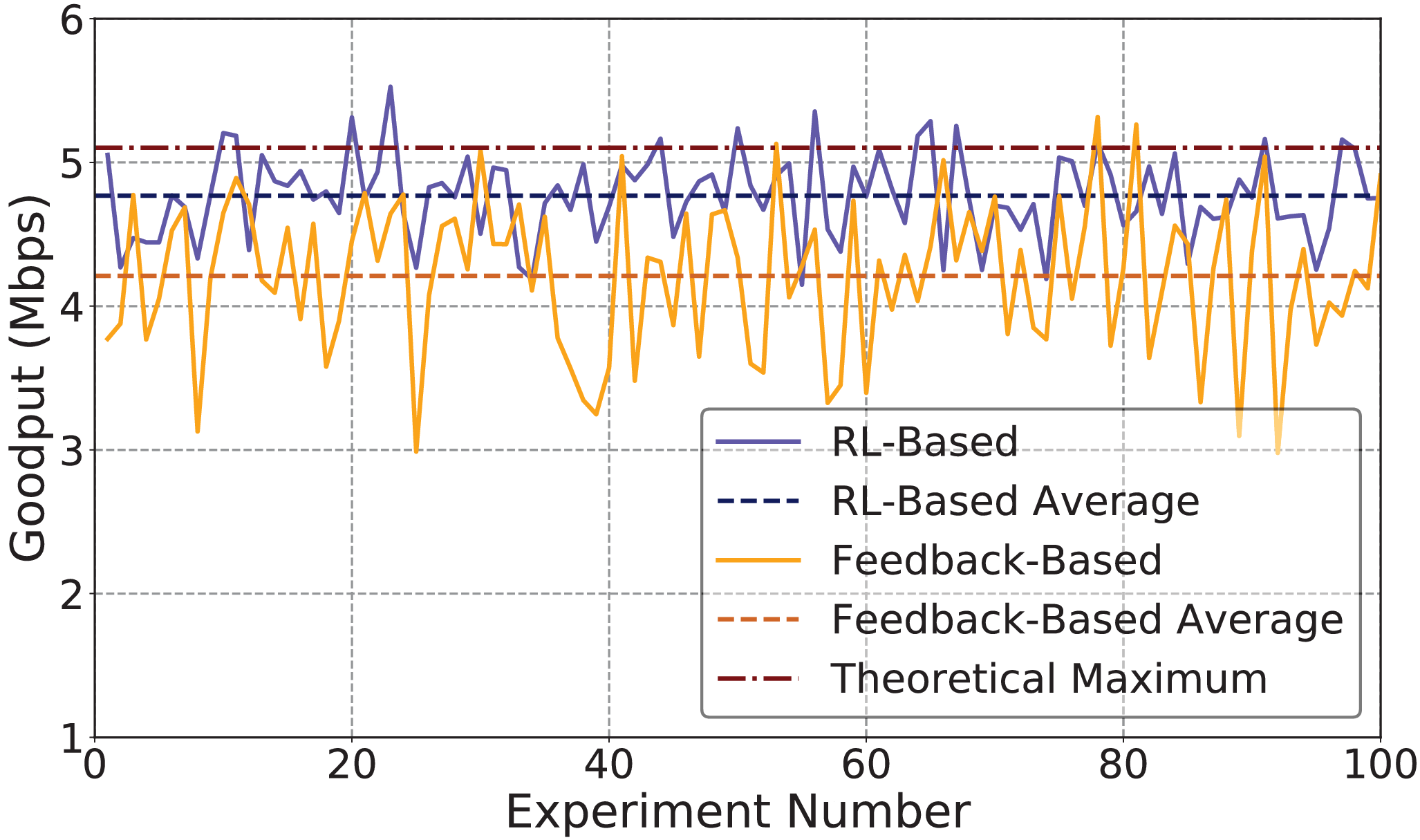}
    \end{subfigure}
    \vfill
    \begin{subfigure}[b]{.48\textwidth}
        \caption{\centering}
        \label{earth-moon-random-5RTT:delivery_delay}
        \vspace{-2ex}
        \includegraphics[width=\textwidth]{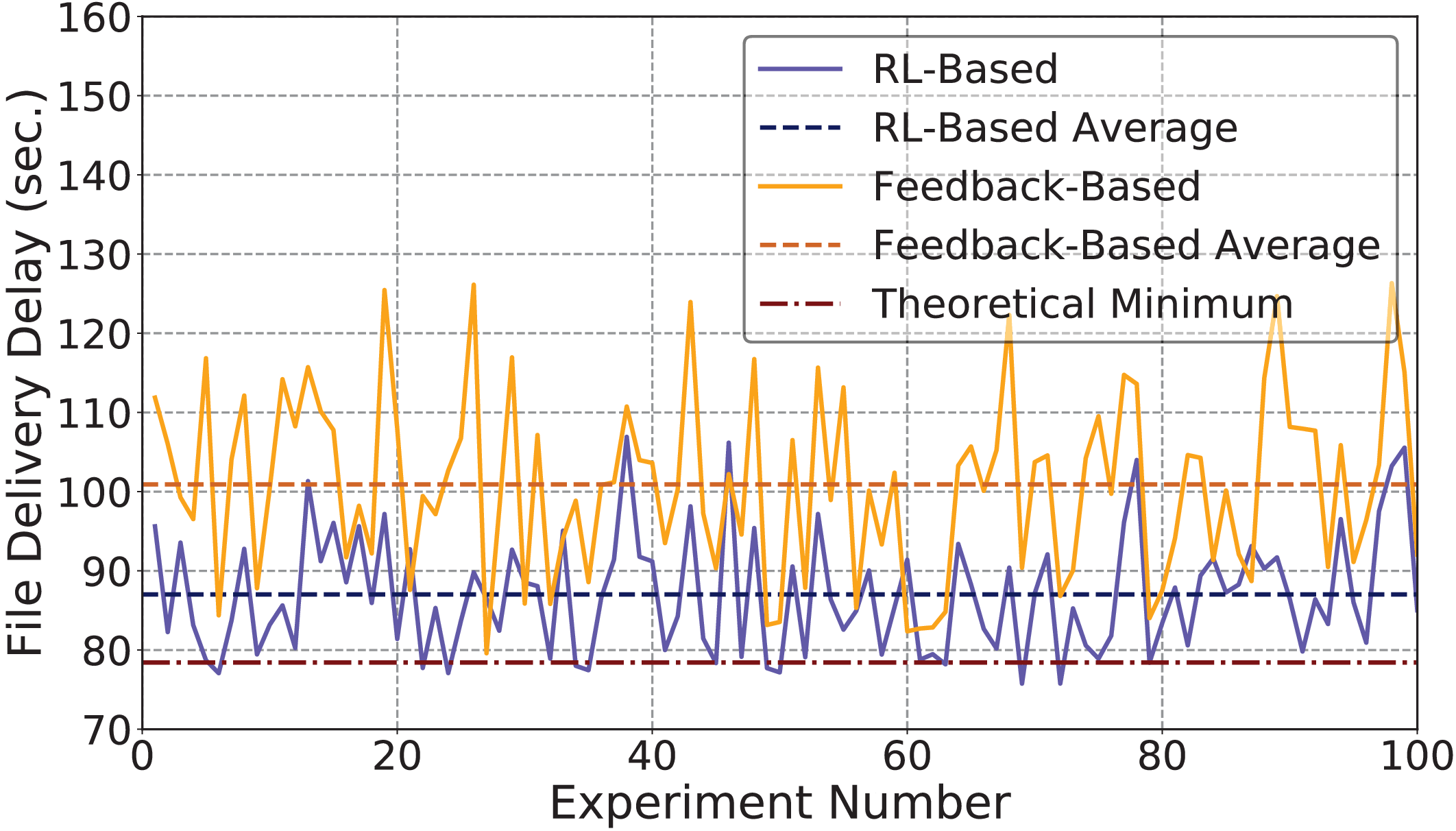}
    \end{subfigure}
    \hfill
    \begin{subfigure}[b]{.48\textwidth}
        \caption{\centering}
        \label{earth-moon-random-10RTT:delivery_delay}
        \vspace{-2ex}
        \includegraphics[width=\textwidth]{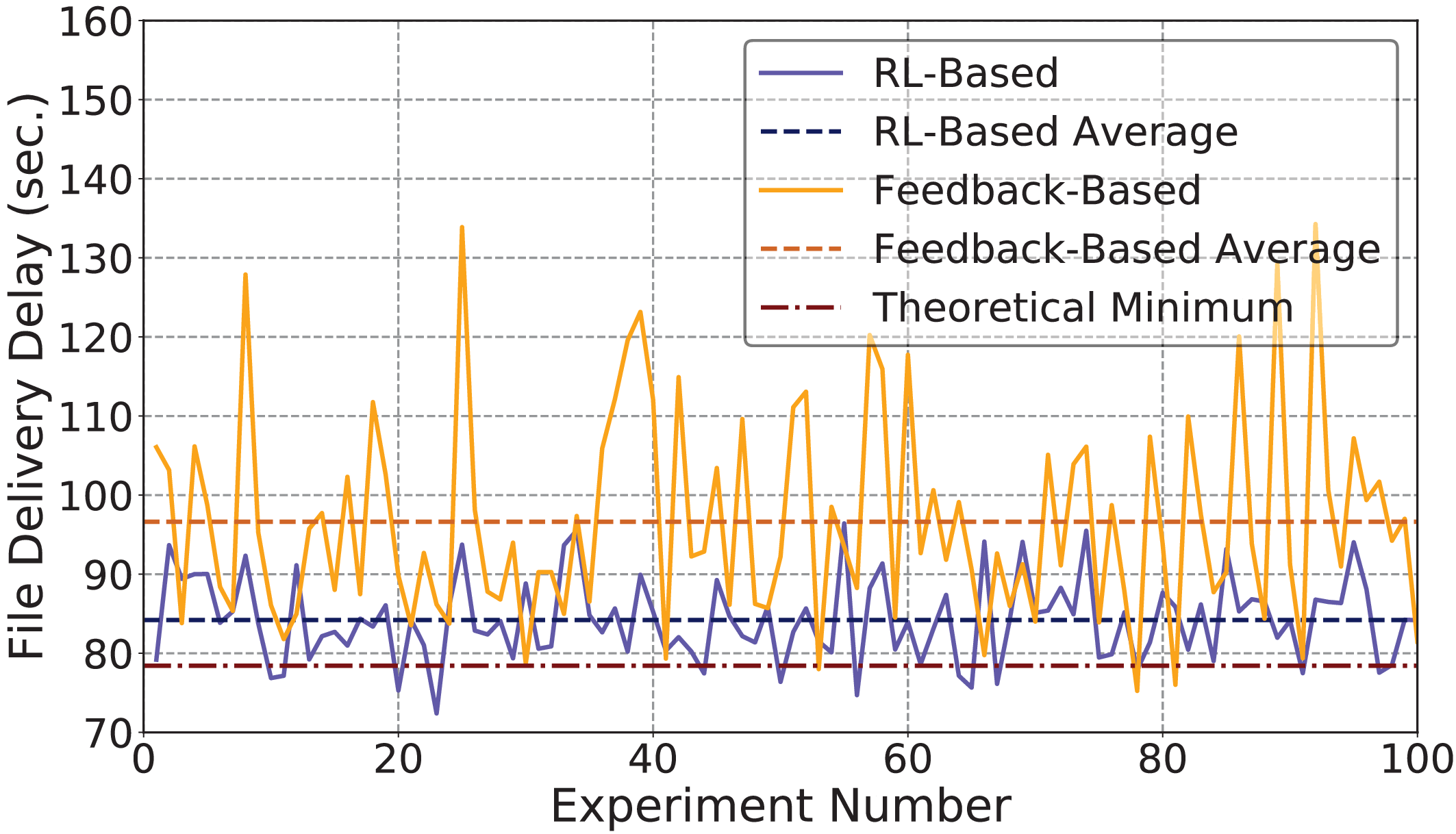}
    \end{subfigure}
    \vfill
    \begin{subfigure}[b]{.48\textwidth}
        \caption{\centering}
        \label{earth-moon-random-5RTT:fail_decoding}
        \vspace{-2ex}
        \includegraphics[width=\textwidth]{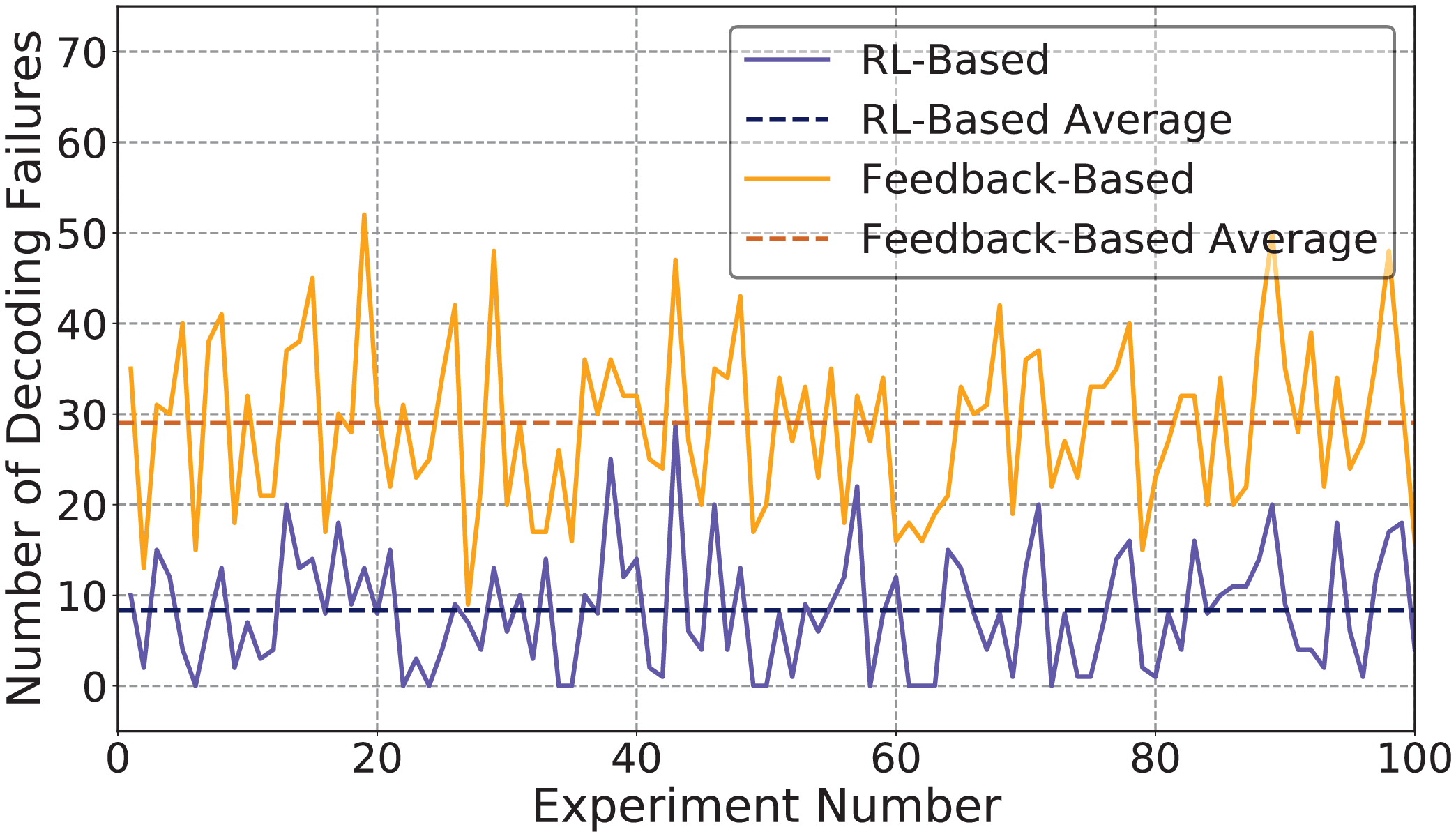}
    \end{subfigure}
    \hfill
    \begin{subfigure}[b]{.48\textwidth}
        \caption{\centering}
        \label{earth-moon-random-10RTT:fail_decoding}
        \vspace{-2ex}
        \includegraphics[width=\textwidth]{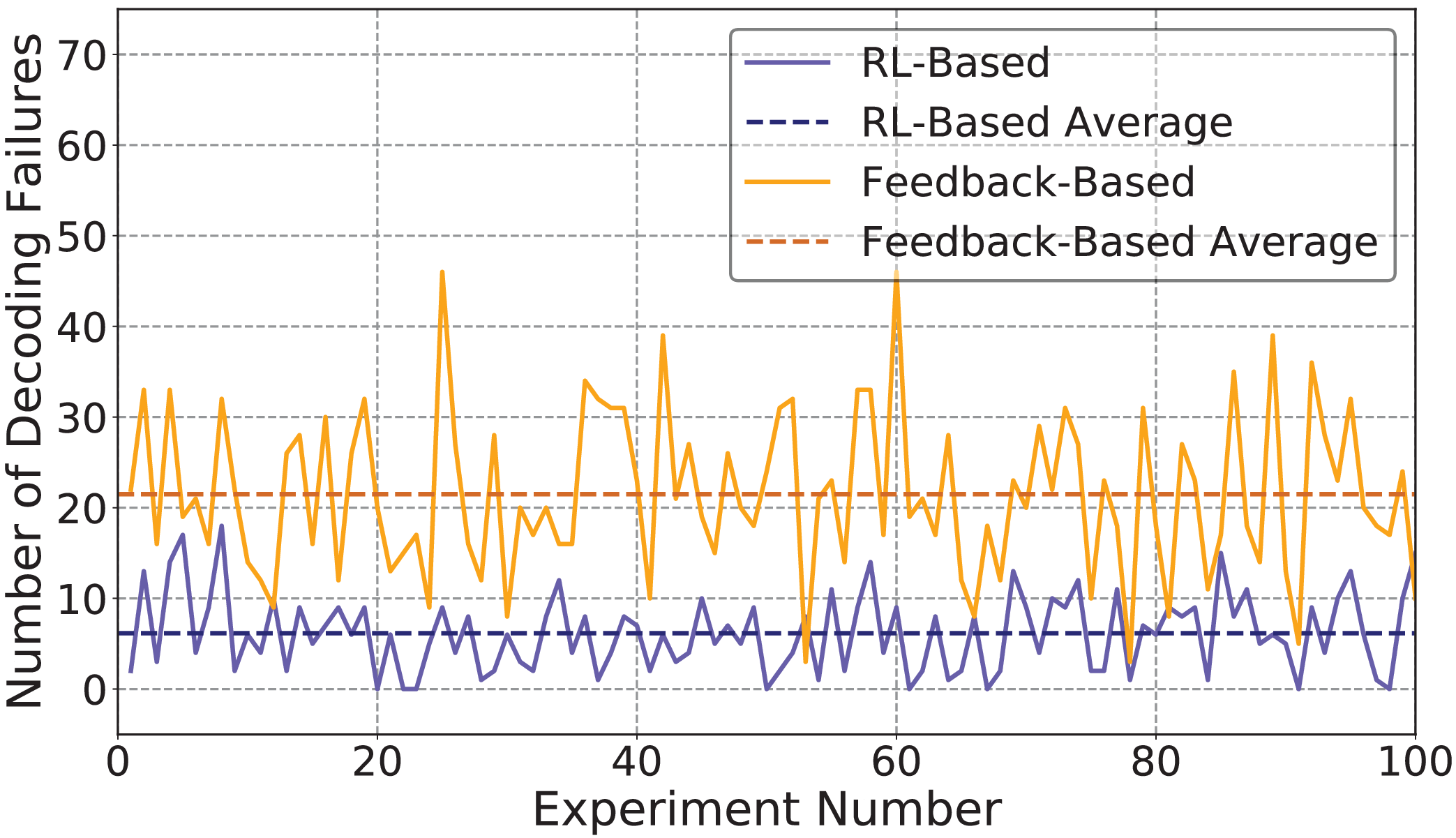}
    \end{subfigure}
    \caption{Performance comparison in the Earth-Moon scenario, with RTT = 2 s and downlink using discrete uniform loss model. (a, b) Goodput; (c, d) File delivery delay; (e, f) Number of matrix decoding failures. Subfigures (a, c, e) correspond to a variation interval of 5 RTTs, while (b, d, f) correspond to 10 RTTs.}
    \label{earth-moon-random}
    \vspace{-15pt}
\end{minipage}
\hfill
\begin{minipage}[t]{0.5\textwidth}
    \centering
    \begin{subfigure}[b]{.48\textwidth}
        \caption{\centering}
        \label{earth-moon-mmc-5RTT:goodput}
        \vspace{-2ex}
        \includegraphics[width=\textwidth]{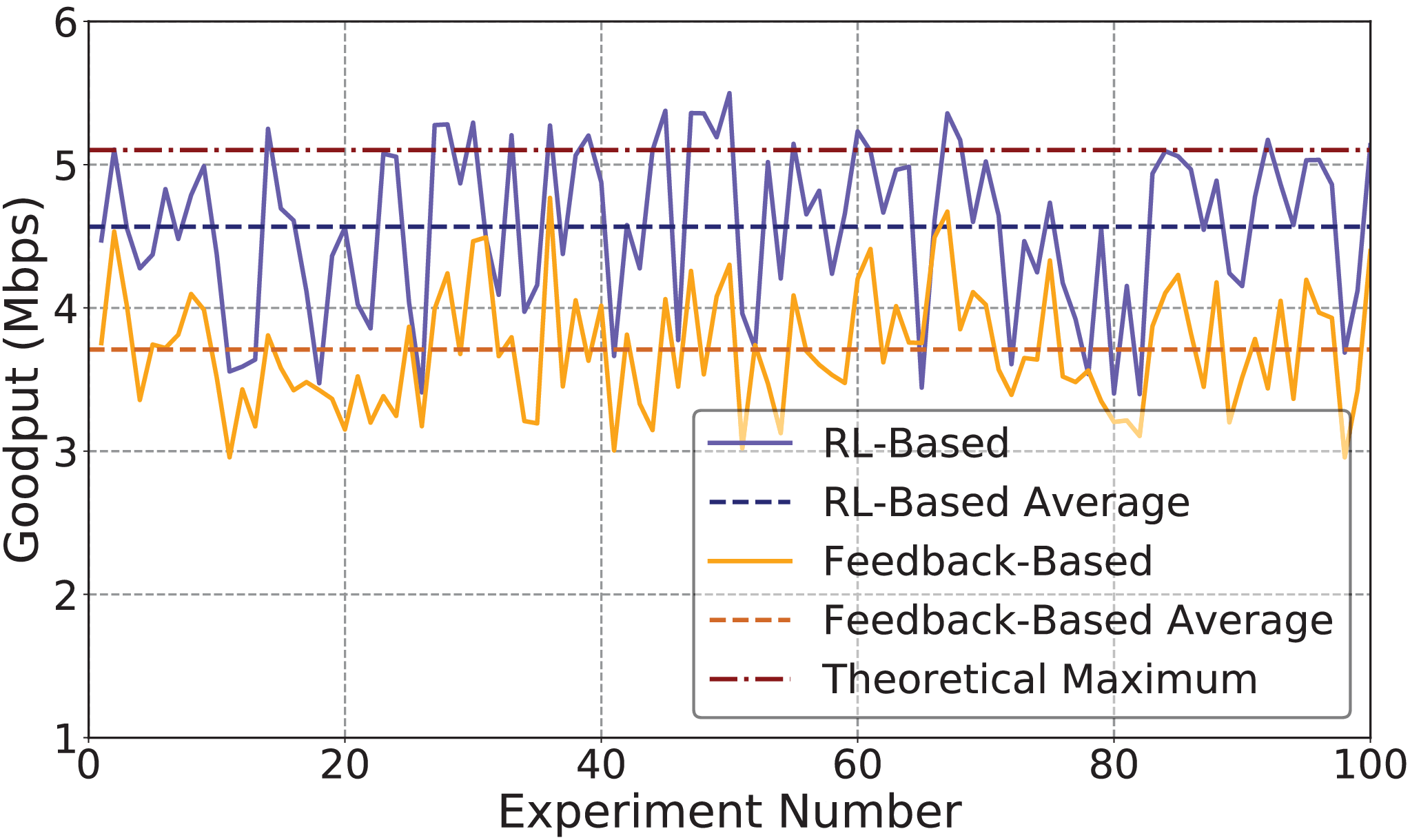}
    \end{subfigure}
    \hfill
    \begin{subfigure}[b]{.48\textwidth}
        \caption{\centering}
        \label{earth-moon-mmc-10RTT:goodput}
        \vspace{-2ex}
        \includegraphics[width=\textwidth]{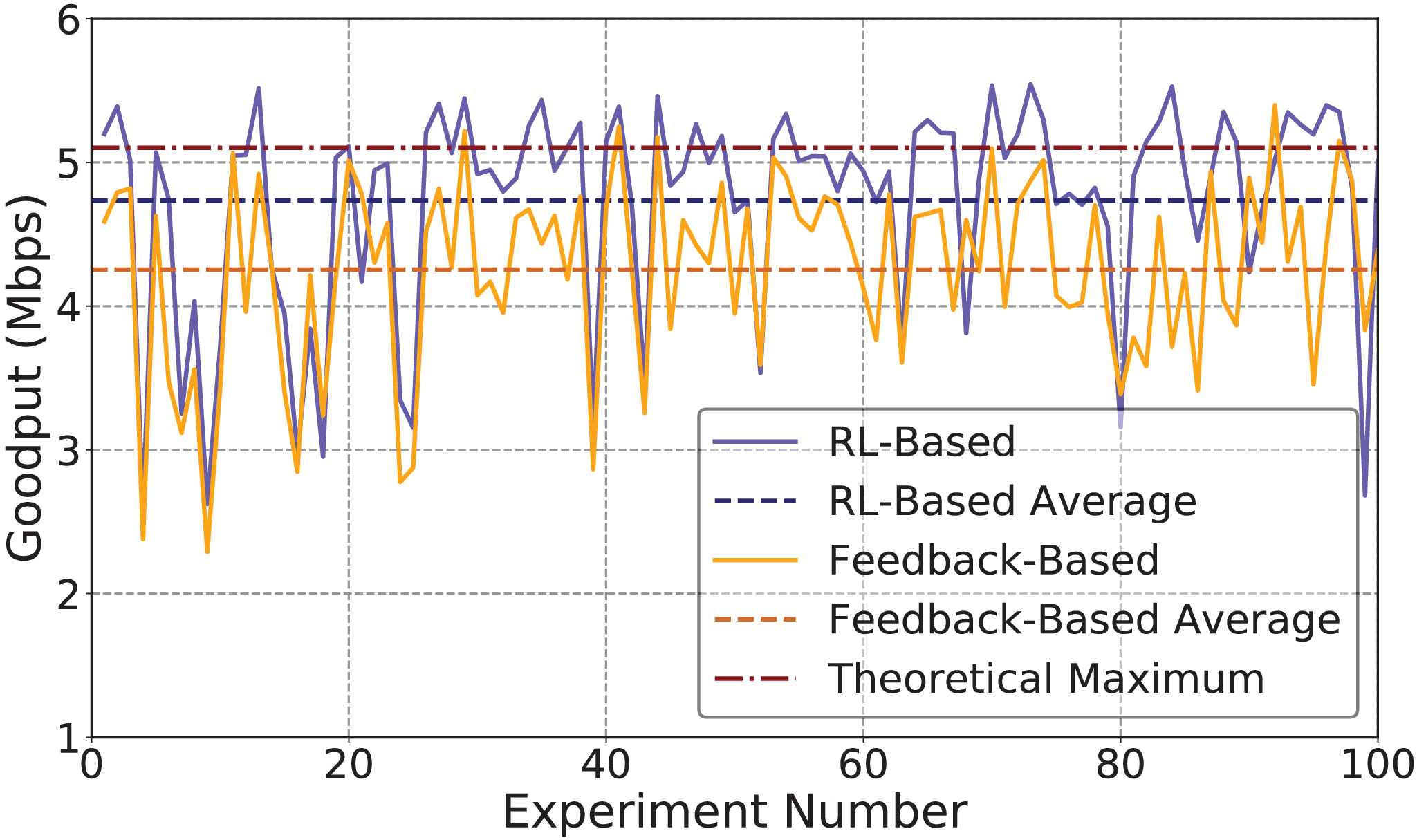}
    \end{subfigure}
    \vfill
    \begin{subfigure}[b]{.48\textwidth}
        \caption{\centering}
        \label{earth-moon-mmc-5RTT:delivery_delay}
        \vspace{-2ex}
        \includegraphics[width=\textwidth]{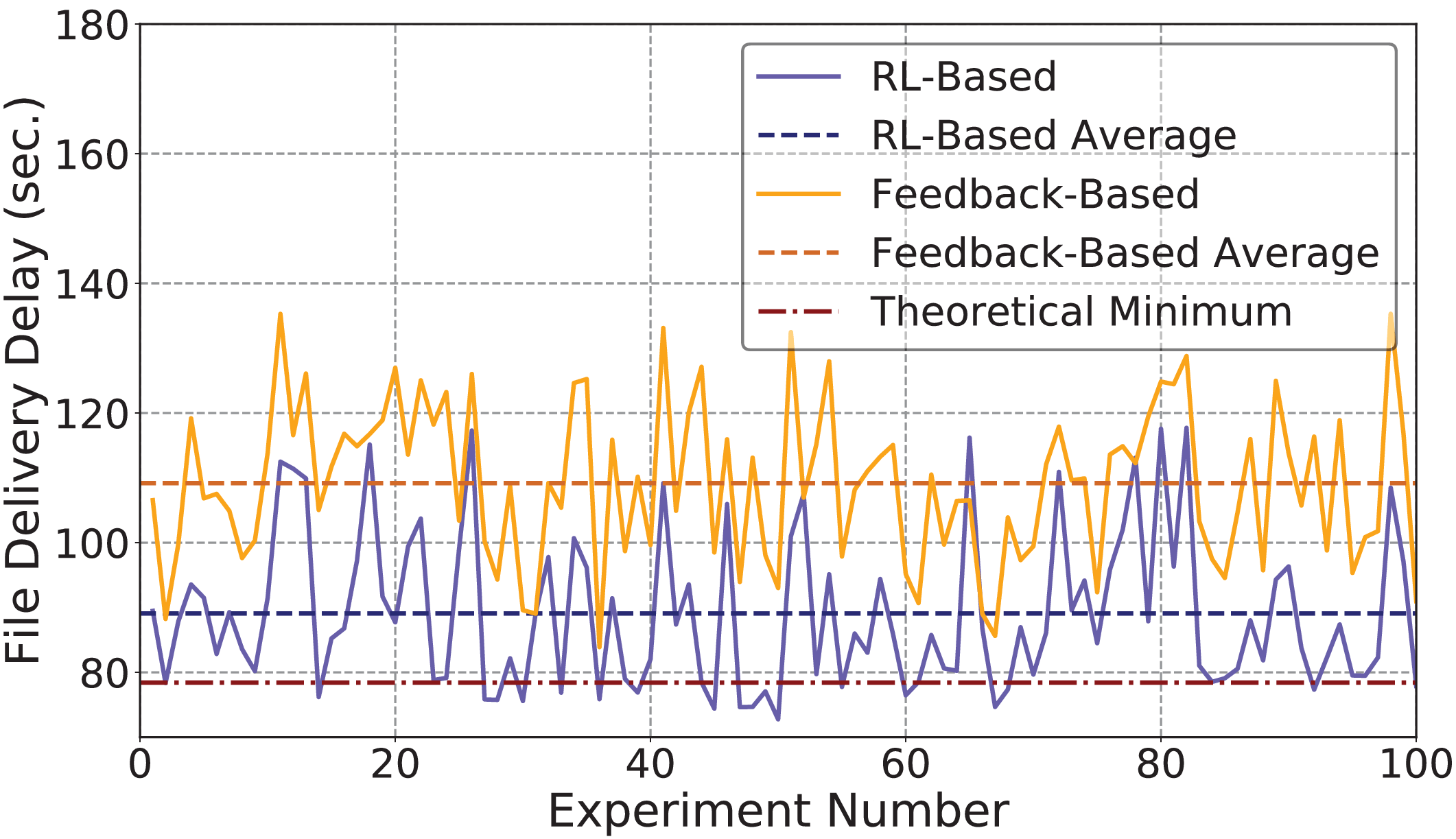}
    \end{subfigure}
    \hfill
    \begin{subfigure}[b]{.48\textwidth}
        \caption{\centering}
        \label{earth-moon-mmc-10RTT:delivery_delay}
        \vspace{-2ex}
        \includegraphics[width=\textwidth]{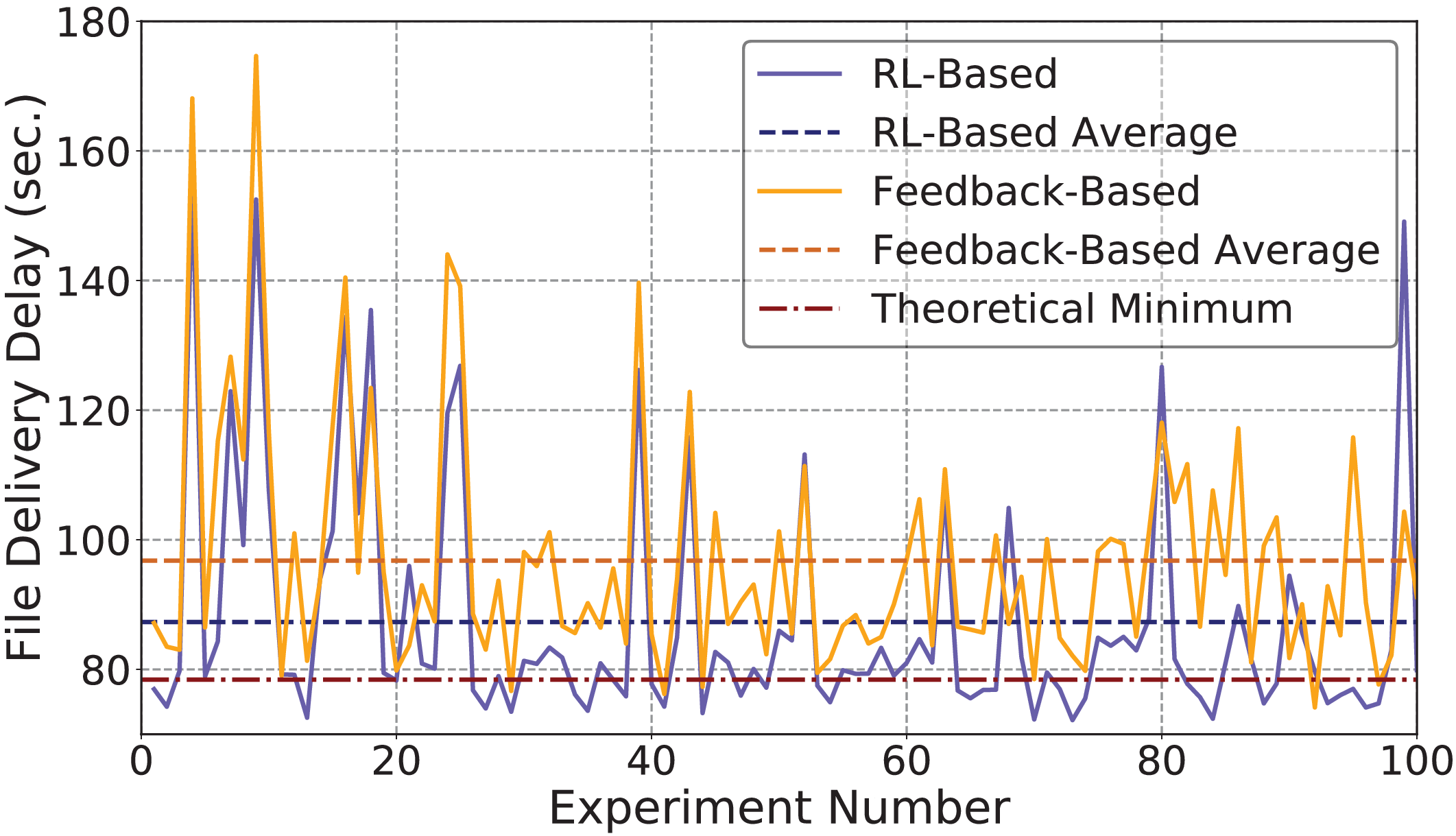}
    \end{subfigure}
    \vfill
    \begin{subfigure}[b]{.48\textwidth}
        \caption{\centering}
        \label{earth-moon-mmc-5RTT:fail_decoding}
        \vspace{-2ex}
        \includegraphics[width=\textwidth]{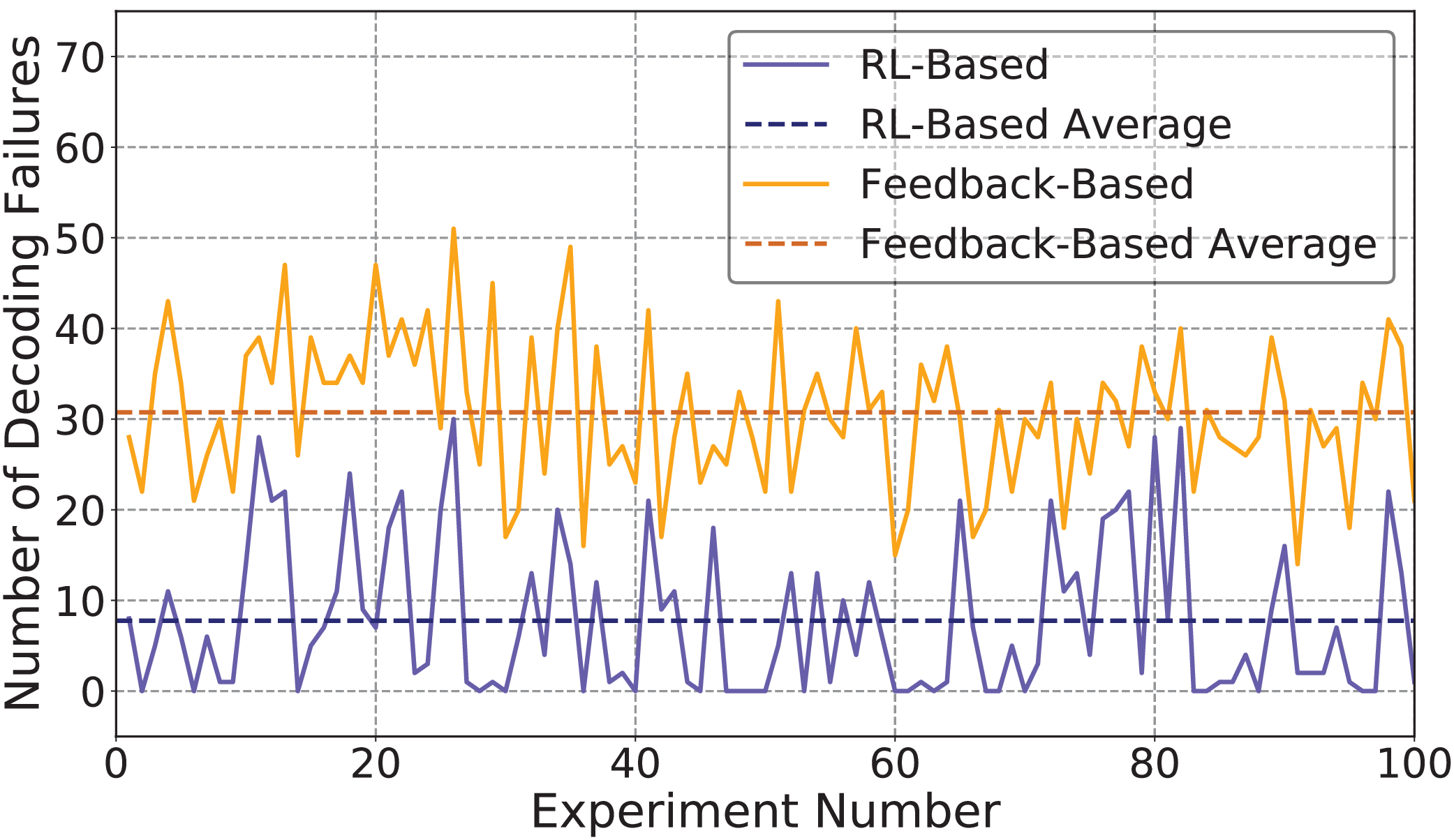}
    \end{subfigure}
    \hfill
    \begin{subfigure}[b]{.48\textwidth}
        \caption{\centering}
        \label{earth-moon-mmc-10RTT:fail_decoding}
        \vspace{-2ex}
        \includegraphics[width=\textwidth]{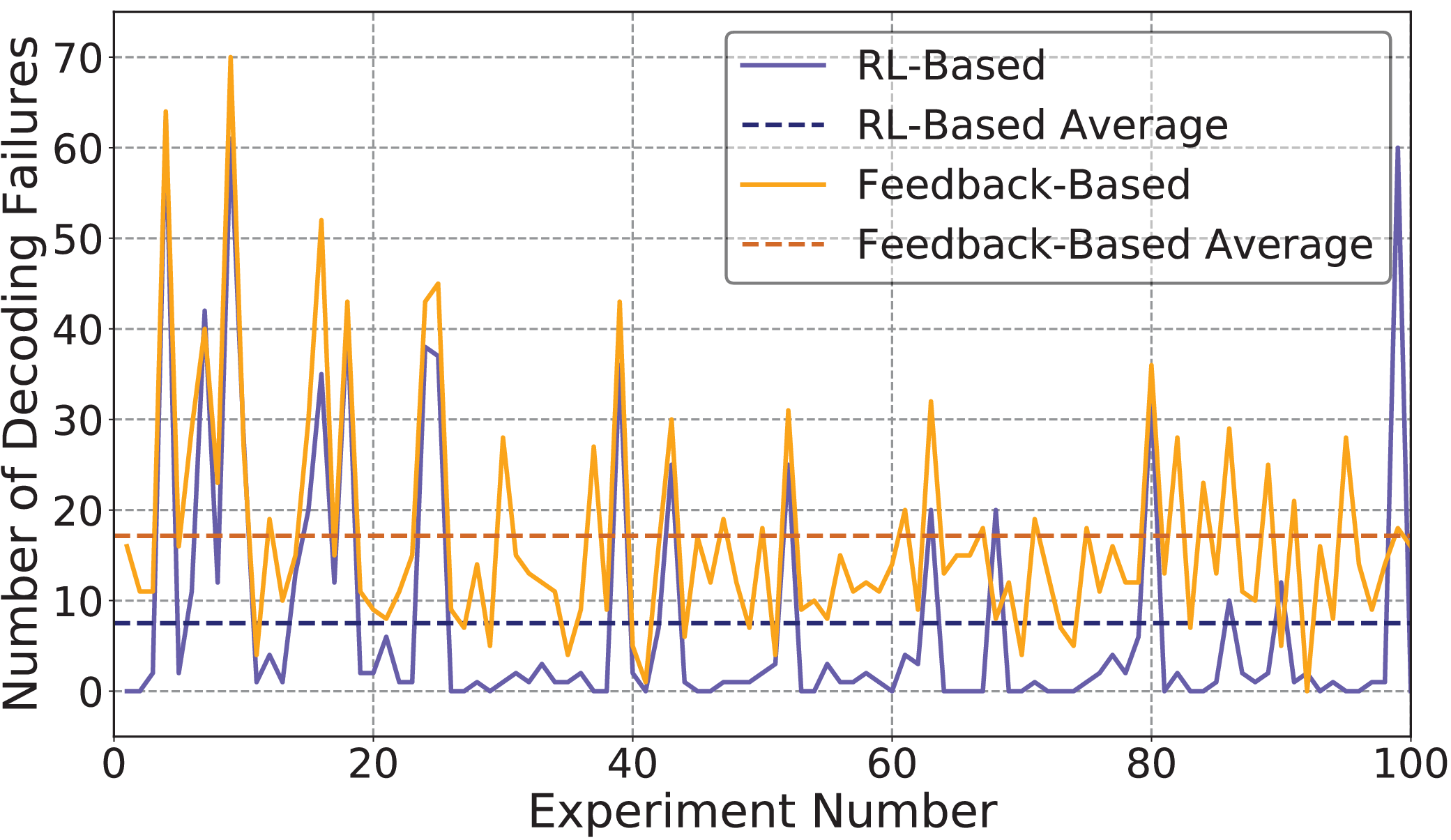}
    \end{subfigure}
    \caption{Performance comparison in the Earth-Moon scenario, with RTT = 2 s and downlink using Markov chain loss model. (a, b) Goodput; (c, d) File delivery delay; (e, f) Number of matrix decoding failures. Subfigures (a, c, e) correspond to an expected variation interval of 5 RTTs, while (b, d, f) correspond to 10 RTTs.}
    \label{earth-moon-mmc}
    \vspace{-15pt}
\end{minipage}
\end{figure*}

\begin{table*}[!t]
\begin{minipage}[t]{0.5\textwidth}
    \centering
    \caption{Performance Comparison (Average Values):\\Earth-Moon Scenario, discrete uniform lossy downlink,\\RTT = 2 s, variation interval = 5 or 10 RTTs.}
    \label{table:comparison_in_earth_moon_random_loss}
    \resizebox{\columnwidth}{!}{
    \begin{tabular}{c|c|c|c|c}
    \hline
    \multicolumn{2}{c|}{\diagbox[height=5em,width=11em]{~~~Scheme}{\\Metric}} & \makecell{Goodput\\(Mbps)} & \makecell{File\\delivery\\delay\\(s)} & \makecell{Number\\of\\decoding\\failures} \\
    \hline
    \hline
    \multirow{2}{*}{\makecell{5\\RTTs}} & RL-Based & \textbf{4.630} & \textbf{87.005} & \textbf{8.340} \\
    \cline{2-5} & Feedback-Based & 4.014 & 100.913 & 29.010 \\
    \hline
    \multirow{2}{*}{\makecell{10\\RTTs}} & RL-Based & \textbf{4.769} & \textbf{84.198} & \textbf{6.160} \\
    \cline{2-5} & Feedback-Based & 4.211 & 96.616 & 21.500 \\
    \hline
    \multicolumn{2}{c|}{Fixed $p_e$ (20\%)} & 5.102 & 78.424 & 0 \\
    \hline
    \end{tabular}
    }
\end{minipage}
\hfill
\begin{minipage}[t]{0.5\textwidth}
    \centering
    \caption{Performance Comparison (Average Values):\\Earth-Moon Scenario, Markov chain lossy downlink,\\RTT = 2 s, expected variation interval = 5 or 10 RTTs.}
    \label{table:comparison_in_earth_moon_mmc_loss}
    \resizebox{\columnwidth}{!}{
    \begin{tabular}{c|c|c|c|c}
    \hline
    \multicolumn{2}{c|}{\diagbox[height=5em,width=11em]{~~~Scheme}{\\Metric}} & \makecell{Goodput\\(Mbps)} & \makecell{File\\delivery\\delay\\(s)} & \makecell{Number\\of\\decoding\\failures} \\
    \hline
    \hline
    \multirow{2}{*}{\makecell{5\\RTTs}} & RL-Based & \textbf{4.567} & \textbf{89.063} & \textbf{7.750} \\
    \cline{2-5} & Feedback-Based & 3.709 & 109.190 & 30.740 \\
    \hline
    \multirow{2}{*}{\makecell{10\\RTTs}} & RL-Based & \textbf{4.735} & \textbf{87.294} & \textbf{7.510} \\
    \cline{2-5} & Feedback-Based & 4.255 & 96.777 & 17.140 \\
    \hline
    \multicolumn{2}{c|}{Fixed $p_e$ (20\%)} & 5.102 & 78.424 & 0 \\
    \hline
    \end{tabular}
    }
\end{minipage}
\end{table*}

\subsection{Earth-Moon Scenario}
In the simulated Earth-Moon communication scenario (RTT = 2 s), we initially applied a a discrete uniform packet loss model to the downlink, with $p_e$ varying at 5 or 10 RTT intervals. Fig.~\ref{earth-moon-random} illustrates the comprehensive performance of both schemes over 100 rounds of file transmission under these conditions, with corresponding average values detailed in Table~\ref{table:comparison_in_earth_moon_random_loss}. Additionally, we conducted supplementary transmission tests to determine the theoretical performance upper bound for all schemes under identical conditions. While maintaining other link parameters constant, we fixed $p_e$ at 20\% (the median of the 0-35\% range) and configured ECLSA to have prior knowledge of this information. The results of this ideal scenario test are also presented in the figures and table.

Due to the complete randomness of link variations, each round of file transmission may have experienced different link conditions, resulting in significant performance fluctuations between rounds. The RL scheme outperformed the ECLSA scheme in approximately 90\% of the file transmission rounds. We attribute the few cases where the RL scheme underperformed to two main reasons: first, the RL agent retains a 1\% probability of random action selection post-training to explore potential new situations, often leading to incorrect rate selections; second, the completely random jumps of $p_e$ may cause the RL's predictive strategy to be incorrect in some cases, underperforming compared to ECLSA's conservative strategy.

However, these instances occurred infrequently, and thus, the introduction of RL significantly enhanced overall performance in most cases. When $p_e$ varied every 5 RTTs (10 s), the RL scheme increased average goodput by 0.626 Mbps and reduced average file delivery delay by 13.908 s. When the variation interval extended to 10 RTTs (20 s), the RL scheme still excelled despite the more gradual link quality changes theoretically favoring the ECLSA scheme. In this case, goodput improved by 0.558 Mbps, and delivery delay decreased by 12.418 s. Analysis of decoding failure statistics reveals that RL significantly reduced matrix decoding failures. This improvement directly led to a substantial decrease in LTP retransmissions. In other words, the RL agent assisted ECLSA-LTP in selecting more appropriate coding rates during transmissions in most cases, thus leading to overall performance improvement.

Notably, the RL scheme still showed a performance gap in both cases compared to the ideal scenario. This is attributed to the persistence of a few decoding failures, which triggered LTP retransmissions, consequently leading to the performance disparity. RL did not eliminate decoding failures for two reasons. Firstly, as previously mentioned, the RL agent occasionally did incorrect actions. Secondly, independent of the $R_c$ control algorithms, ECLSA-LTP's minimum $R_c$ of 2/3 was insufficient to perfectly handle scenarios where $p_e$ was 35\%, resulting in some matrices still failing to decode due to excessive symbol loss.

While the above experiments have validated the effectiveness of the RL-FEC algorithm in time-varying links, some unrealistic assumptions remain. In actual space environments, link quality varies continuously due to various factors. However, the current model controls $p_e$ through random jumps, which diverges from reality. Moreover, the fixed time intervals for $p_e$ changes represent an idealized simplification.

Based on these considerations, we conducted new experiments using the Markov chain packet loss model for the downlink to demonstrate the generalizability and practicality of RL-FEC further. The link modeling in this experiment should be more complex: a Markov Chain replaced the Discrete Uniform Distribution, and temporal uncertainty was introduced. The expected intervals for $p_e$ changes were set to 5 or 10 RTTs. Fig.~\ref{earth-moon-mmc} and Table~\ref{table:comparison_in_earth_moon_mmc_loss} illustrate the experimental results for both schemes under these conditions. The RL scheme achieved 4.567 Mbps, 89.063 s, 7.750 decoding failures, and 4.735 Mbps, 87.294 s, and 7.510 decoding failures in the two cases, respectively. On average, these results demonstrate that even in more realistic and complex simulated links, the performance of our proposed RL-based adaptive FEC-LTP algorithm did not show significant degradation compared to the discrete uniform loss scenario. The RL agent could still learn about the transmission task and link variations without prior knowledge, optimizing the expected performance metrics.

The performance data for the ECLSA scheme were 3.709 Mbps, 109.190 s, 30.740 decoding failures, and 4.255 Mbps, 96.777 s, and 17.140 decoding failures, showing a substantial gap compared to the RL scheme. When the variation interval extended to 10 RTTs, the ECLSA scheme narrowed its performance gap with the RL scheme, corroborating that this change is more favorable for ECLSA scheme.

As previously mentioned, at a 35\% $p_e$, neither scheme could prevent some matrix decoding failures. Consequently, their corresponding data curves in Fig.~\ref{earth-moon-mmc} occasionally exhibit simultaneous significant fluctuations, for example, during the 1th to 20th transmission experiment rounds. Due to the longer RTT in the simulated Earth-Mars scenario, which could potentially lead to excessively long durations for individual file transmission experiments, we removed 35\% from the set of $p_e$ values for subsequent experiments in this scenario.

\subsection{Earth-Mars Scenario}

Earth-Mars communication presents more formidable challenges compared to Earth-Moon communication. Firstly, signals traversing this extensive interplanetary link encounter a more complex and variable space environment. Secondly, the highly long RTT between Earth and Mars exacerbates the negative impact of LTP's retransmission mechanism on communication performance. This characteristic significantly alters the existing trade-off between reliability and efficiency. In the Earth-Mars scenario, the primary objective of FEC $R_c$ control inevitably shifts towards more conservative strategies to ensure reliable transmission rather than focusing on reducing coding redundancy to improve transmission efficiency.

Interestingly, these extreme communication conditions may simplify the RL scheme's task of finding optimal strategies. Compared to the Earth-Moon scenario, the decision space in Earth-Mars communication may be more distinct, enabling the RL scheme to identify and converge on effective optimization strategies more easily.

Given the long RTT (4 min) in the simulated Earth-Mars communication scenario, experiments under additional conditions were deemed time-consuming and unnecessary, considering the break in the trade-off mentioned above. Finally, we conducted experiments in a Markov chain lossy link, with an expected $p_e$ variation interval of 5 RTTs. As in the Earth-Moon scenario, we performed a performance test with a fixed $p_e$ of 20\% and informed ECLSA in advance without modifying other link conditions. This test represents the ideal performance upper bound for all schemes.

\afterpage{
\begin{figure}[!t]
    \centering
    \begin{subfigure}[b]{.48\textwidth}
        \caption{\centering}
        \label{earth-mars-mmc-5RTT:goodput}
        \vspace{-2ex}
        \includegraphics[width=\textwidth]{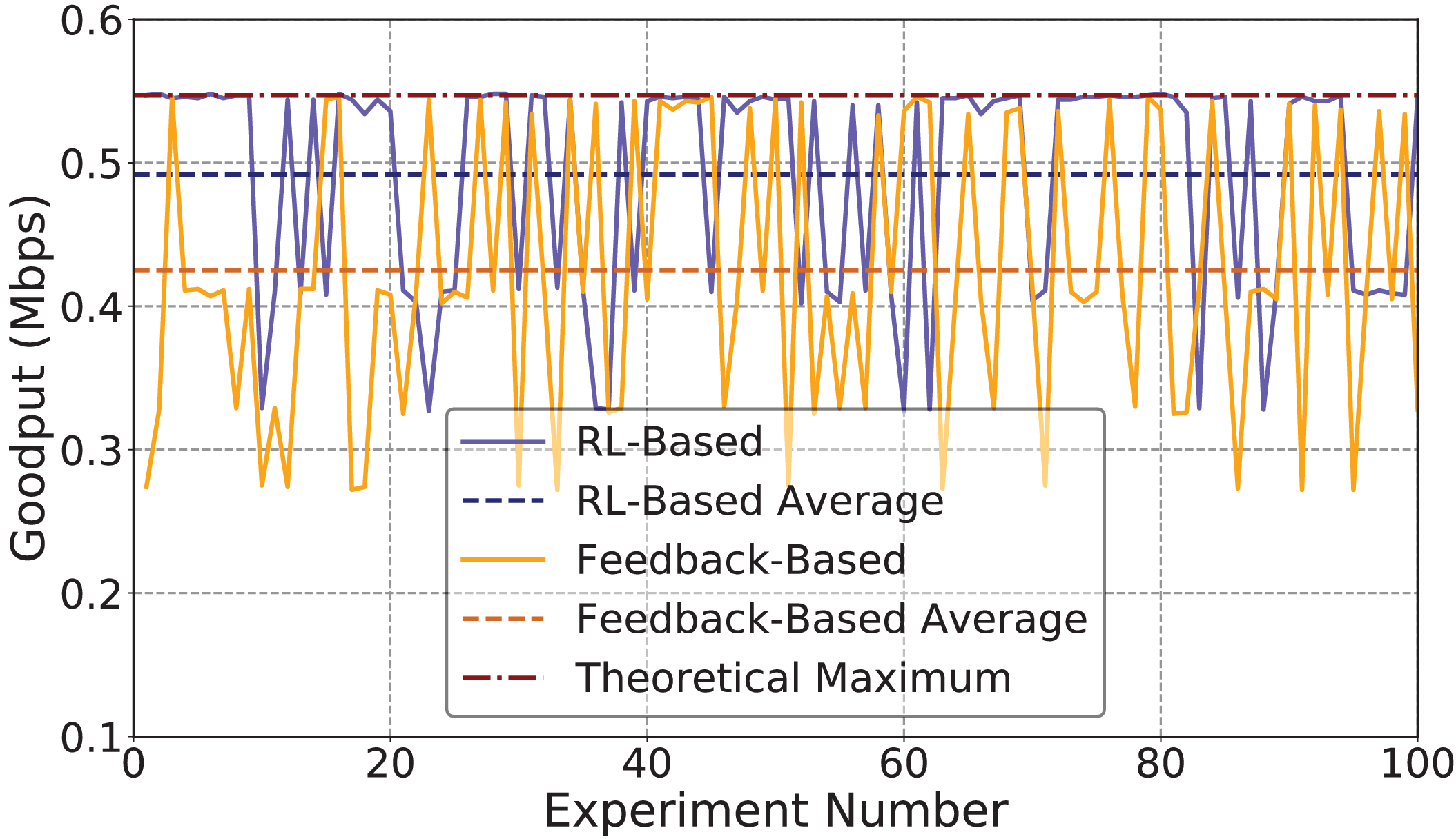}
    \end{subfigure}
    \vfill
    \begin{subfigure}[b]{.48\textwidth}
        \caption{\centering}
        \label{earth-mars-mmc-5RTT:delivery_delay}
        \vspace{-2ex}
        \includegraphics[width=\textwidth]{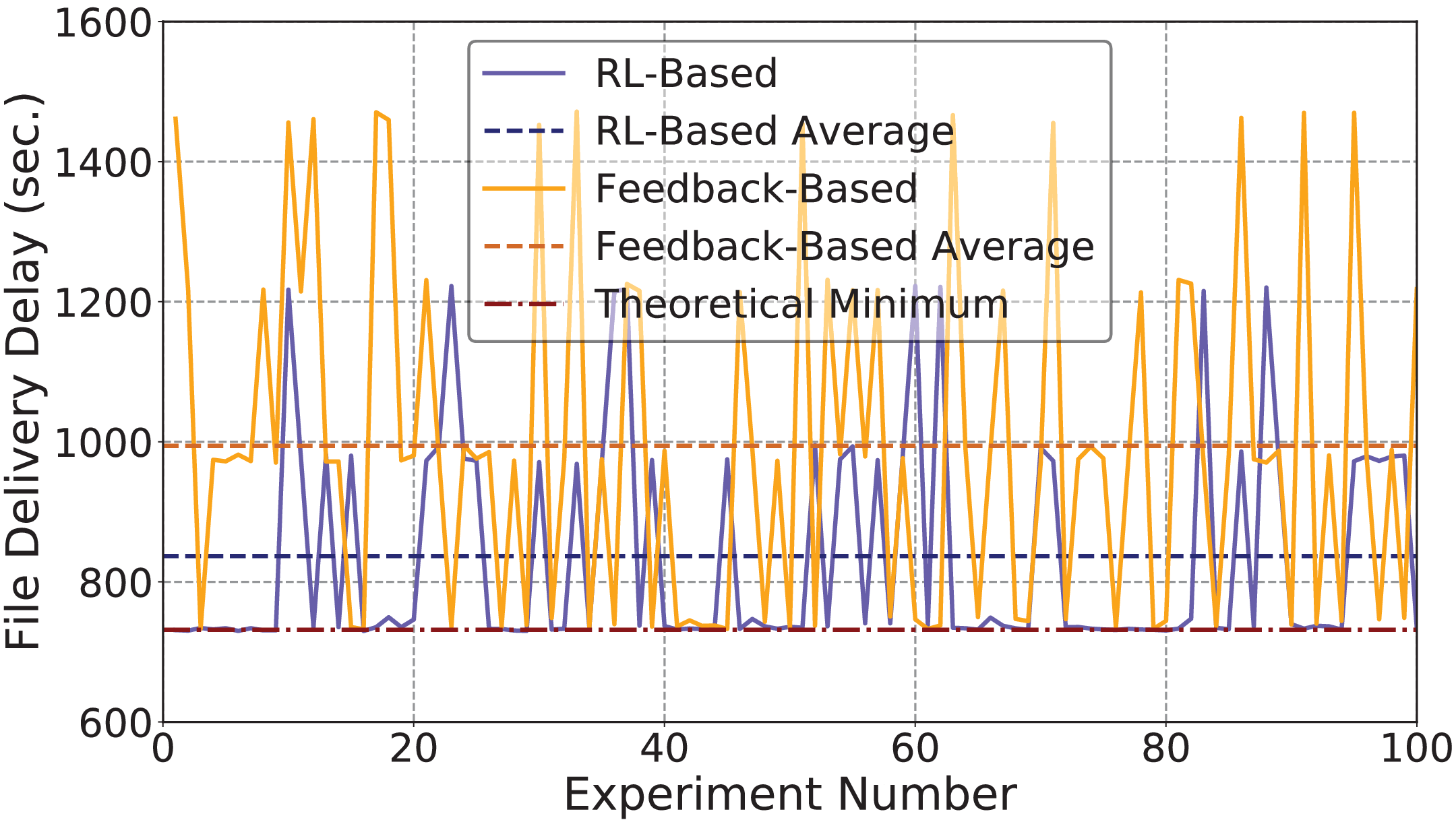}
    \end{subfigure}
    \vfill
    \begin{subfigure}[b]{.48\textwidth}
        \caption{\centering}
        \label{earth-mars-mmc-5RTT:fail_decoding}
        \vspace{-2ex}
        \includegraphics[width=\textwidth]{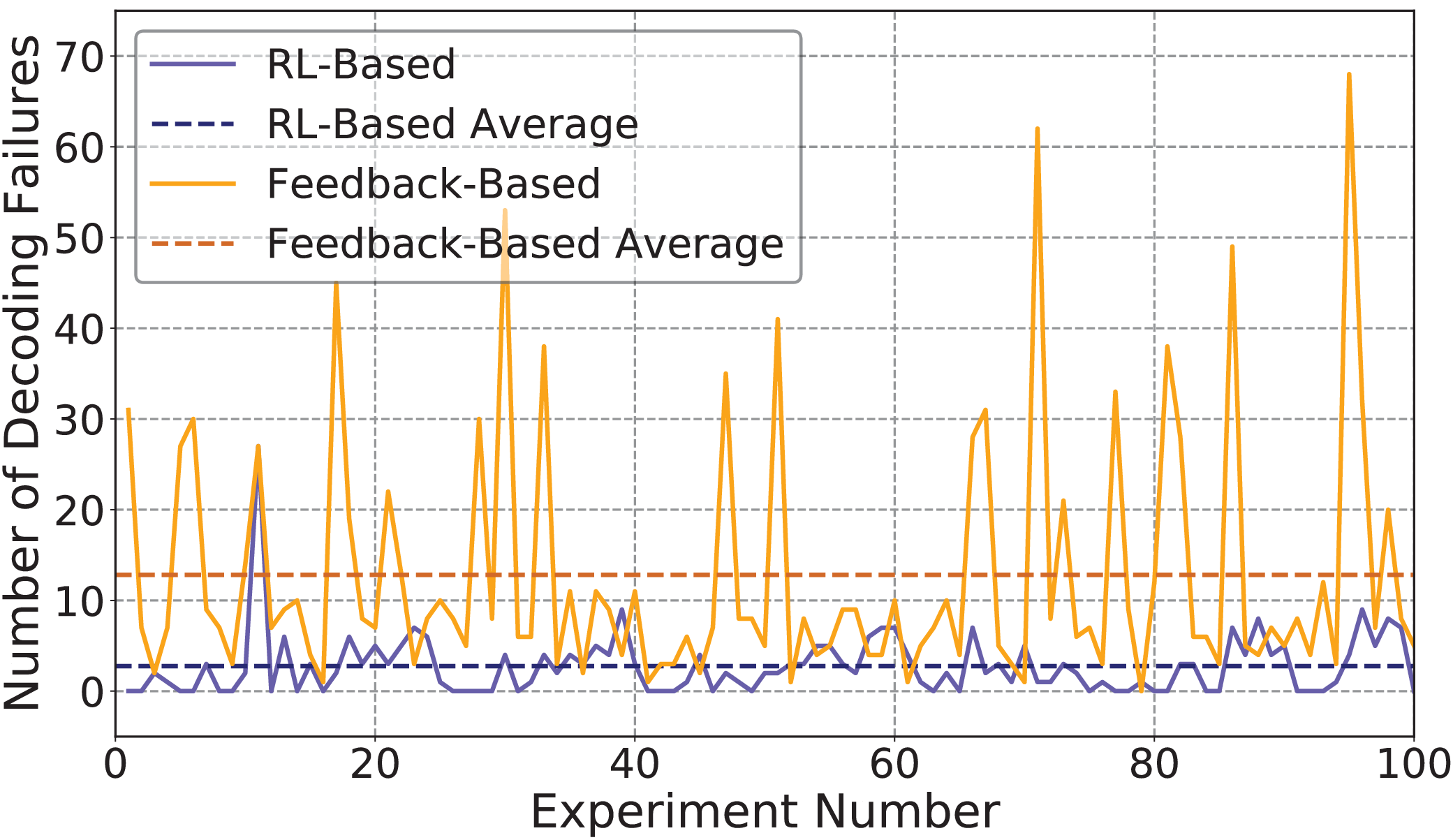}
    \end{subfigure}
    \caption{Performance comparison in the Earth-Mars scenario, with RTT = 4 min and downlink using Markov chain loss model. The expected variation interval is 5 RTTs, i.e., 20 min. (a) Goodput; (b) File delivery delay; (c) Number of matrix decoding failures.}
    \label{earth-mars-mmc}
    \vspace{-10pt}
\end{figure}

\begin{table}[!t]
\centering
\caption{Performance Comparison (Average Values):\\Earth-Mars Scenario, Markov chain lossy downlink,\\RTT = 4 min, expected variation interval = 5 RTTs.}
\label{table:comparison_in_earth_mars_mmc_loss}
\resizebox{\columnwidth}{!}{%
\begin{tabular}{c|c|c|c}
\hline
\diagbox[height=5em,width=8em]{Scheme}{\\Metric} & \makecell{Goodput\\(Mbps)} & \makecell{File\\delivery\\delay\\(s)} & \makecell{Number\\of\\decoding\\failures} \\
\hline
\hline
RL-Based & \textbf{0.492} & \textbf{836.920} & \textbf{2.760} \\
\hline
Feedback-Based & 0.425 & 994.231 & 12.820 \\
\hline
Fixed $p_e$ (20\%) & 0.547 & 731.583 & 0 \\
\hline
\end{tabular}%
}
\end{table}
}

Fig.~\ref{earth-mars-mmc} illustrates the performance comparison between the RL scheme and the ECLSA scheme over 100 rounds of file transmission under these conditions, with the corresponding average values presented in Table~\ref{table:comparison_in_earth_mars_mmc_loss}. Due to the long RTT and the limited number of parallel LTP sessions, the RL scheme and the ECLSA scheme, as illustrated in Fig.~\ref{earth-mars-mmc}\subref{earth-mars-mmc-5RTT:goodput}, did not exhibit a significant difference in goodput. The RL scheme achieved an average goodput of 0.492 Mbps, while the ECLSA scheme reached 0.425 Mbps. If we focus more on file delivery delay, a coupling parameter with goodput, we will discover a surprising reduction. The RL scheme achieved an average delivery delay of 836.72 s, significantly lower than the ECLSA scheme's 994.231 s. This improvement translated to a time saving of approximately 2.5 minutes per 50 MB file transmitted under identical conditions. However, compared to the ideal scenario's 731.583 s, a gap of about 1.5 minutes remained. Similar to the experiments in the Earth-Moon scenario, this phenomenon can be explained by analyzing the statistics of decoding failures. Compared to the ECLSA scheme, the RL scheme reduced the average number of decoding failures by nearly 10 instances, ultimately achieving an average of 2.760 decoding failures. While this improvement was substantial, the remaining decoding failures were the primary reason for the average delivery delay exceeding that of the ideal scenario.

As mentioned earlier, with 35\% removed from the set of $p_e$ values, decoding failures were nearly eliminated in some transmission rounds. This resulted in goodput and delivery delay approaching ideal performance. Due to the RL scheme's superior ability to eliminate decoding failures, this phenomenon was more pronounced and frequent in this approach.

\section{Conclusion}
In this paper, we designed, analyzed, implemented, and comprehensively evaluated an RL-based adaptive FEC-LTP scheme for delayed feedback links. The primary conclusion is that a well-designed RL scheme can effectively control FEC adaptive coding transmission over lossy interplanetary links with long propagation delays, improving goodput and reducing file delivery latency. This work demonstrates the potential of applying RL to interplanetary communications. Future research will focus on expanding the role of RL in DTN protocols, considering additional optimization objectives to enhance further the overall performance of DTN protocols in interplanetary links.

\ifCLASSOPTIONcaptionsoff
  \newpage
\fi

\printbibliography


\enlargethispage{-5in}

\end{document}